\documentclass[%
amsmath,amssymb,
aps,
jcp,reprint,floatfix
]{revtex4-2}

\usepackage[utf8]{inputenc}
\usepackage[version=4]{mhchem}
\newcommand{\mc}{\multicolumn}
\usepackage{bbm}
\usepackage{dcolumn}
\usepackage{mathtools}
\usepackage{enumitem}
\usepackage{letltxmacro}
\LetLtxMacro{\ORIGselectlanguage}{\selectlanguage}
\makeatletter
\DeclareRobustCommand{\selectlanguage}[1]{%
  \@ifundefined{alias@\string#1}
    {\ORIGselectlanguage{#1}}
    {\begingroup\edef\x{\endgroup
       \noexpand\ORIGselectlanguage{\@nameuse{alias@#1}}}\x}%
}
\newcommand{\definelanguagealias}[2]{%
  \@namedef{alias@#1}{#2}%
}
\makeatother
\usepackage{physics}
\usepackage{{booktabs}}
\usepackage[unicode]{hyperref}
\hypersetup{
   unicode=true,          
   plainpages=false,
   colorlinks=true,       
   linkcolor=black,          
   citecolor=black,        
   filecolor=black,      
   urlcolor=blue 
}

\usepackage{orcidlink}
\DeclareUnicodeCharacter{2212}{-}
\DeclareUnicodeCharacter{2009}{\,}
\newcommand{\etal}{\emph{et al.}}

\newcommand{\rref}[1]{Eq.\ (\ref{#1})}

\newcommand{\bd}[1]{\textbf{#1}} 
\newcommand{\lrr}[1]{\left({#1}\right)} 
\newcommand{\bA}{{\bf A}}

\newcommand{\br}{{\bf r}}

\newcommand{\ur}[1]{\mathrm{#1}}

\newcommand{\nr}{\addtocounter{equation}{1}\tag{\theequation}} 

\newcommand{\imax}{\alpha_{\ur{max}}^{(i)}}
\newcommand{\smax}{\alpha_{\ur{max}}^{(s)}}
\newcommand{\pmax}{\alpha_{\ur{max}}^{(p)}}
\newcommand{\iD}{D^{(i)}}
\newcommand{\sD}{D^{(s)}}
\newcommand{\pD}{D^{(p)}}
\renewcommand{\theequation}{\arabic{section}.\arabic{equation}}


\begin{abstract}
Despite the fact that most quantum chemistry basis sets are designed for accurately modelling valence chemistry, these general-purpose basis sets continue to be widely used to model core-dependent properties. Core-specialised basis sets are designed with specific features to accurately represent the behaviour of the core region. This design typically incorporates Gaussian primitives with higher exponents to capture core behaviour effectively, as well as some decontraction of basis functions to provide flexibility in describing the core electronic wave function. The highest Gaussian exponent and the degree of contraction for both $s$- and $p$-basis functions effectively characterise these design aspects. 

In this study, we compare the design and performance of general-purpose basis sets against several literature basis sets specifically designed for three core-dependent properties: J coupling constants, hyperfine coupling constants, and magnetic shielding constants (used for calculating chemical shifts). Our findings consistently demonstrate a significant reduction in error when employing core-specialised basis sets, often at a marginal increase in computational cost compared to the popular 6-31G** basis set. Notably, for expedient calculations of J coupling, hyperfine coupling and magnetic shielding constants, we recommend the use of the pcJ-1, EPR-II, and pcSseg-1, basis sets respectively. For higher accuracy, the pcJ-2, EPR-III, and pcSseg-2 basis sets are recommended.

\end{abstract}

\begin{document}

\title{On the Specialisation of Gaussian Basis Sets for Core-Dependent Properties}
\author{Robbie T. Ireland \orcidlink{0000-0002-4282-7810}}
\author{Laura K. McKemmish \orcidlink{0000-0003-1039-2143}}
 \email{l.mckemmish@unsw.edu.au}
 \affiliation{ School of Chemistry, UNSW Sydney, Australia}

\maketitle

\section{Introduction} \label{sec: Intro}

Computational quantum chemistry is ubiquitous and increasingly essential in modern chemistry and related fields \cite{JensenIntroduction2007}. Indeed, decades of research have established well-tested methodologies to develop approximations to the Schr\"{o}dinger equation (e.g. density functional approximations).  These can yield computationally-tractable, yet reasonably-accurate, results when used in conjunction with optimised all-Gaussian basis sets which are themselves used to describe electron distributions \cite{Boys1950Electronic, Atkins2013}. 

Implicit in the framework of most modern computational quantum chemistry packages is that the basis sets typically utilised (6-31G*, cc-pVnZ, def2, pcseg-n and so on) rely on the fact that much of chemistry, including reaction energies and molecular geometries, depends primarily on the behaviour of valence electrons. {However}, the treatment of core electrons by many commonly used basis sets is extremely poor for two reasons. First, the Gaussian primitive is inherently ill-suited for describing the near-nuclei (core) region and in particular completely fails to capture the electron-nuclear cusp \cite{Kato1957}. Second, most general-purpose basis sets are designed without any significant flexibility to describe variation in core electron distribution as a result of differing molecular environments. Fortunately, for most properties, these shortcomings are irrelevant and high accuracy predictions for properties are achievable.

However, the poor treatment of core electrons is a crucial shortcoming in predicting core-dependent properties, most notably nuclear magnetic resonance (NMR) spin-spin (J) coupling constants \cite{Ramsey1953, Helgaker2000Analytical} and NMR shielding constants (used to calculate chemical shifts) \cite{Vaara2007, Helgaker1999}. Other properties dependent on the description of core electrons include hyperfine coupling constants (used throughout electron paramagnetic resonance studies) \cite{Weltner1989,Lindgren1982,Woodgate1983}, isomer shift and quadrupole splittings in Mössbauer spectroscopy \cite{Mossbauer1958, Gutlich1978, Neese2002,Casassa2016}, field shifts when considering finite-sized nuclei \cite{Rosenthal1932,Hanle1978,Ehrenfest1922}, X-ray emission and absorption spectroscopies (XES and XAS, respectively) \cite{Norman2018}, relativistic correction terms such as the mass-velocity and Darwin corrections \cite{Pachucki2005,Jeszenszki2021a, Ireland2021b} and total energies \cite{McKemmish2012}. 

There are a variety of approaches that have been considered to more accurately treat core electrons (summarized in Section \ref{sec:Alternitives}). However, the simplest approach is to design all-Gaussian basis sets specifically for core properties, adding primitives to improve accuracy and additional contracted basis functions to enable basis set flexibility in the core region; this approach is explained more in Sub-Section \ref{subsec:Specialised}. The use of these specialised all-Gaussian basis sets is the approach this paper focuses on reviewing and evaluating. 

{This paper considers the problem of selecting an appropriate basis set for core-dependent properties by quantifying the performance of different basis set choices.   In particular, we quantify the accuracy and timing for core-specialised basis sets against a number of general-purpose basis sets.} We consider three important core-dependent properties: NMR J coupling constants (Section \ref{Sec:IsoJCoupling}), hyperfine coupling constants (Section \ref{Sec:IsoHyperfine}) and NMR shielding constants (Section \ref{sec:IsoShieldingConstants}).

This paper is arranged as follows: 
\begin{itemize}
    \item Section \ref{sec:BasisSetDesign} starts by considering Gaussian basis functions from the lens of core-dependent properties before describing the design of specialised all-Gaussian basis sets.
    \item Section \ref{sec:Methodology} establishes the methodology for our benchmark calculations and rationalises the basis sets to be investigated, level of theory selections, our molecular benchmark data set, computational details and analysis approach. Of note in this section is the first concise summary of, to our knowledge, a large selection of available all-Gaussian core-specialised basis sets specialised towards the calculation of J coupling constants (J-specialised basis sets), hyperfine coupling constants (H-specialised basis sets) and magnetic shielding constants (S-specialised basis sets). 
    \item Three core-dependent properties are considered in depth in Section \ref{Sec:IsoJCoupling} (J-couplings), Section \ref{Sec:IsoHyperfine} (hyperfine couplings) and Section \ref{sec:IsoShieldingConstants} (magnetic shielding constants, i.e. chemical shifts). Each section mathematically describes the operators involved in calculating each observable property and the demands of this property on the design of basis sets specialised towards its calculation. Each section concludes with the relative performance of multiple general-purpose and specialised basis sets in terms of both accuracy and relative timings. Though we consider each property separately, it is worth noting that some operators involved (which are each associated with specific demands on the basis set) appear in more than one property; e.g. the Fermi-contact term appears in both J and hyperfine coupling tensors while spin-orbit terms occur in all three operators. 
    \item Section \ref{sec:Recommendations} provides a concise tabular reference and associated discussion for readers on the double-zeta and triple-zeta basis sets recommended for the three core-dependent properties studied, {with all findings summarised in Section \ref{Sec:FinalRemarks}}. 

\end{itemize}

\section{Basis Set Design} \label{sec:BasisSetDesign}

\subsection{General-Purpose all-Gaussian Basis Set Design}
The basis functions within basis sets are used to describe electron distributions and are thus central to the accuracy of all computational quantum chemistry calculations \cite{Szabo1996, Suzuki1965, Here1986}. Appropriate selection of basis functions is critical for ensuring that electron distributions can be described to a given accuracy compactly, allowing for fast computations. Naively, one might evaluate the quality of a basis set by how well it represents the true (complete basis set) electron distribution for target systems, but in fact many chemically relevant properties are more quickly predicted by making sure the regions of interest, i.e. valence orbitals, are accurately and flexibly described while doing `well-enough' in regions with little effect on most chemical properties, i.e. core orbitals. Flexibility means that a basis set is able to describe the differences between electron distribution in similar molecules and is usually quantified by zeta, the number of contracted basis functions used to describe each valence orbital in a basis set. 

The design of basis sets for valence chemistry has been well reviewed \cite{Jensen2013,hill2013gaussian,nagy2017basis} and we refer the reader to these articles for detailed discussion of topics such as the importance of polarisation functions, augmented (diffuse) functions, the differences between different basis set families and the method of optimisation for basis set composition and parameterisation. Instead, we focus here on basis set design considerations for core electrons. 

\begin{figure}[b]
    \centering
    \includegraphics[scale=0.17]{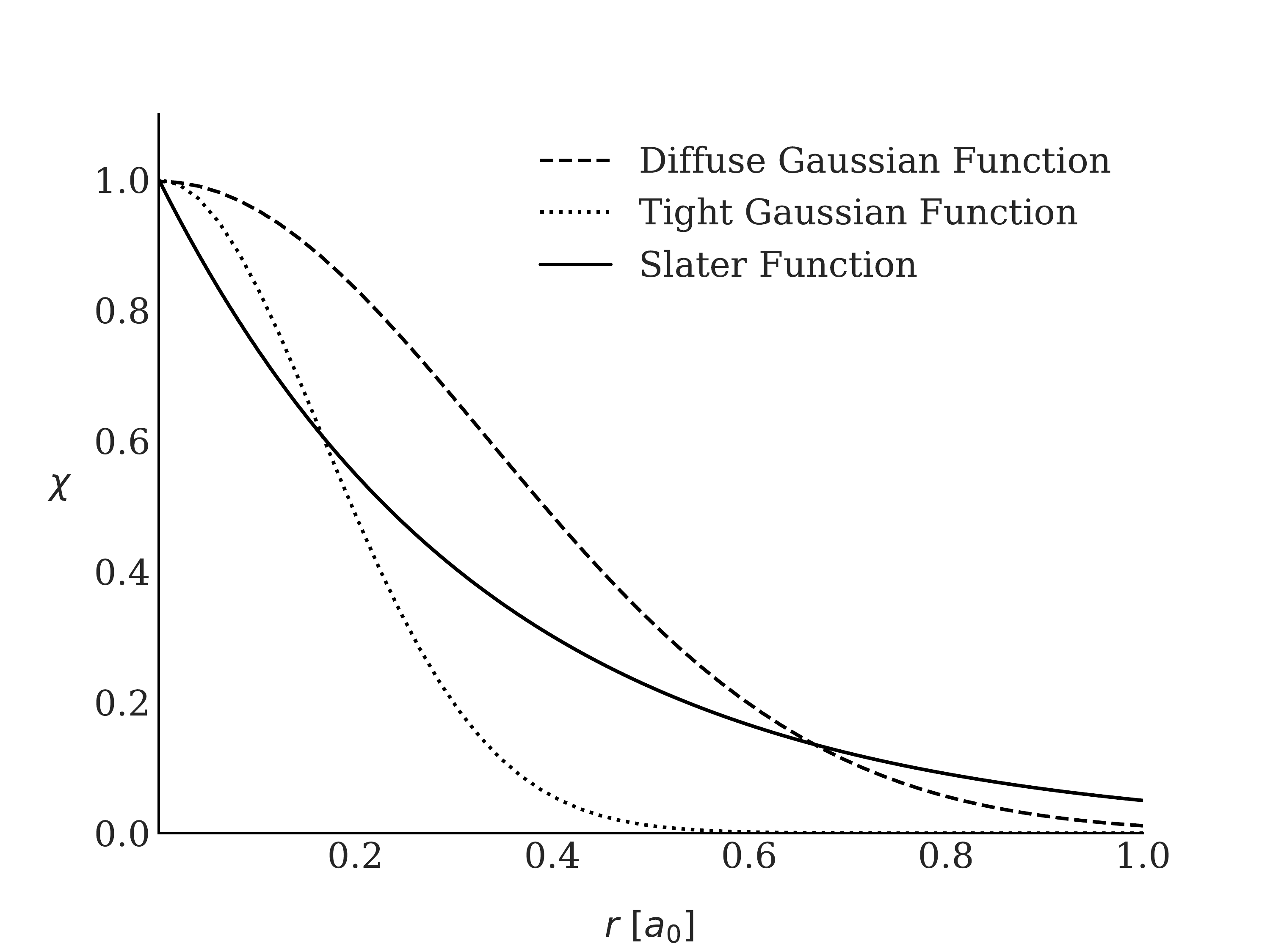}
    \caption{Comparison of the functional form of Slater, diffuse Gaussian and tight Gaussian functions.}
    \label{fig:GausSTO}
\end{figure}

The most important consideration for core electrons is the functional form of the radial functions used in the basis set i.e. the Gaussian functions given by $e^{-\alpha r^2}$. Gaussian basis sets \cite{Boys1950Electronic} have become ubiquitous in modern computational quantum chemistry because the Gaussian product rule allows very efficient analytic evaluation of two-electron integrals \cite{Yarkony1995}. However, Gaussian basis functions completely fail in describing the core behaviour of wave functions. Most notably, Gaussian basis sets are smooth at the point of electron-nuclear coalescence, as shown in Figure \ref{fig:GausSTO}. As such, they cannot satisfy the Kato cusp condition \cite{Kato1957} for the electron-nuclear cusp which is correctly visualised by the Slater function in the same Figure.  

To achieve even a satisfactory representation of electron density near the nuclei, Gaussian basis sets must use a large number of Gaussian primitives (e.g. `6' primitives in the 6-31G* basis set) with high exponents. Even with relatively large numbers of primitives, the electron density at the nucleus is severely under-estimated by general-purpose all-Gaussian basis sets \cite{mckemmish2015accurate}. In particular, the Fermi-contact term which dominates isotropic J coupling and hyperfine coupling is typically very poorly represented. Specialised basis sets are thus necessary for these sorts of core-dependent properties.

\subsection{Specialised all-Gaussian Basis Sets for Core Properties} \label{subsec:Specialised}

Specialised basis sets are customised in design to allow accurate prediction of a specific property. This is a non-trivial task that is most important for properties with unusual basis set demands, such as those that rely on accurate description of electrons in the core (near-nuclei) region.

Generally, specialised basis sets use a `parent' general-purpose set as a template, with several modifications being made to increase accuracy when calculating a particular property. We will refer to such changes made to a general-purpose basis as `specialisation' of the basis set. Multiple authors have specialised basis sets for specific properties, many of which are summarised in Table \ref{table:doubezetabasis}. To the best of our knowledge, however, there is no concise review of core-specialisation overall; this forms one of the motivations for this work.

Different properties will probe different parts (spaces) of the molecular wave function, meaning the calculation of different properties will make different demands of a basis set. Thus, specialisation of a basis set is aimed at improving the description of a particular  functional space(s) ($s$-space, $p$-space, etc...).  The descriptions of such functional spaces are primarily defined by functions of the same angular momentum quantum number $l$ ($s$-functions for $s$-spaces, for example).

A specialised basis set will inherit the description of the different functional spaces from its underlying parent basis set. Then, the alterations made will aim to improve the description of the relevant functional space(s). 
 When generating basis sets specialised towards core-dependent properties, two types of alterations are usually made; including more high-exponent (tight) functions; and limiting the amount of contraction of core functions.   

Tight functions can be added to improve the description of the basis set around the nucleus. This can be seen in Fig. \ref{fig:GausSTO}, where the tight Gaussian behaves more-so like the Slater function around the $r=0$ region than the diffuse Gaussian. Note that one could also remove `redundant' functions - those inherited from the parent basis set which are no longer needed in the specialised basis set - in order to reduce computational demand. 

Contraction of a basis set is often employed as a time-saving measure in typical quantum chemistry calculations, but comes at the cost of accuracy via a reduction in the flexibility of the basis. For example, if we consider a contraction scheme such as 14$s$7$p$ $\to$ 9$s$4$p$ (number of primitive Gaussians $\to$ number of contracted Gaussians), then the quality of the description of the $s$- and $p$-spaces are reduced in the contracted basis set compared to the uncontracted basis set. Thus, we may wish to decontract some or all of the core functions when specialising towards a core-dependent property in order to increase the flexibility of the set in the core region.  

\begin{figure*}[htbp!]
    \centering
    \includegraphics[scale=0.20]{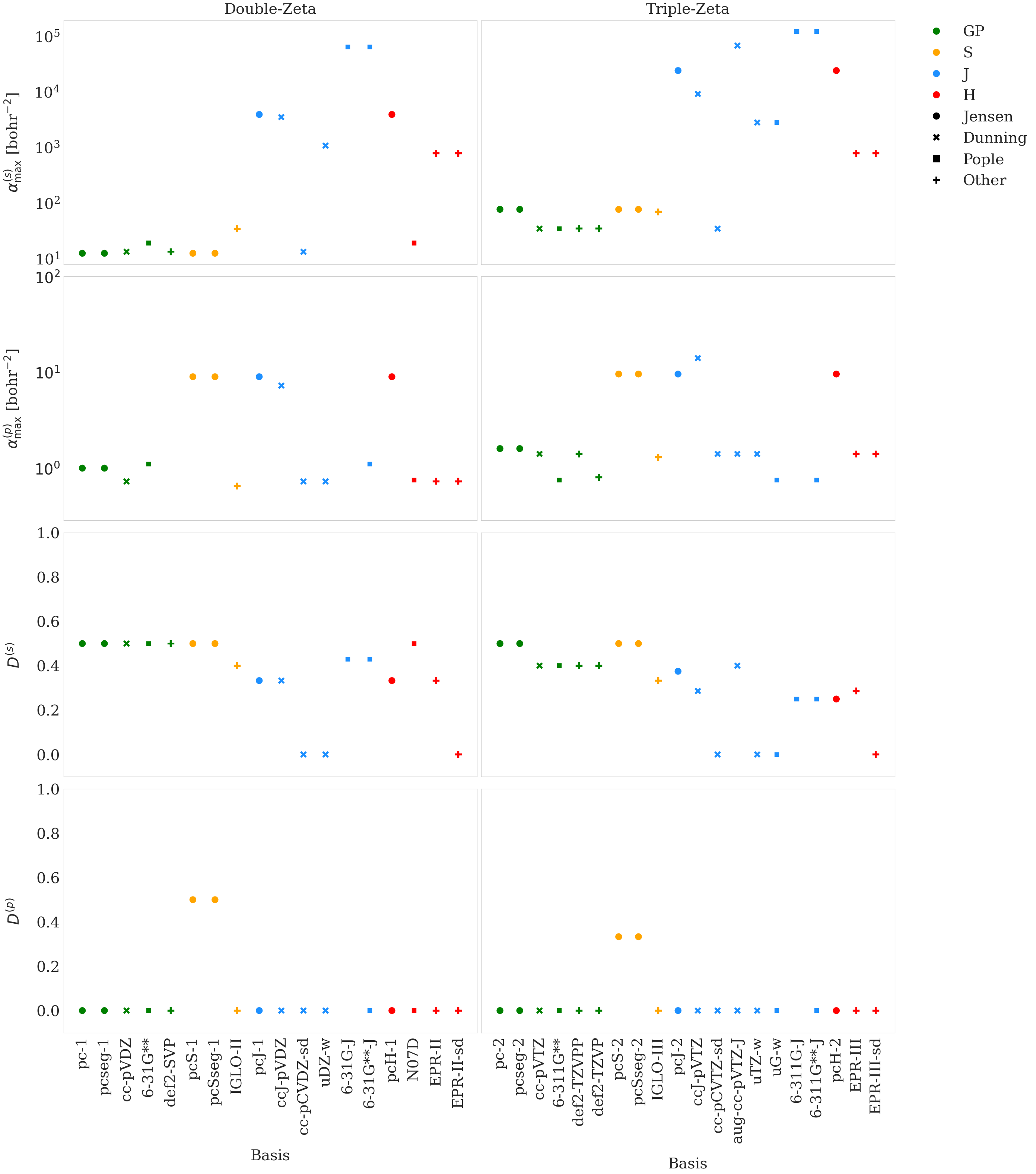}
    \caption{Largest exponent values and degrees of contraction for $s$- and $p$-functions of various double- and triple-zeta basis sets describing the hydrogen atom. Note that the 6-31G-J basis set is unpolarised and thus does not contain any $p$-functions.}
    \label{fig:HydrogenFigs}
\end{figure*}

\begin{figure*}[htbp!]
    \centering
    \includegraphics[scale=0.2]{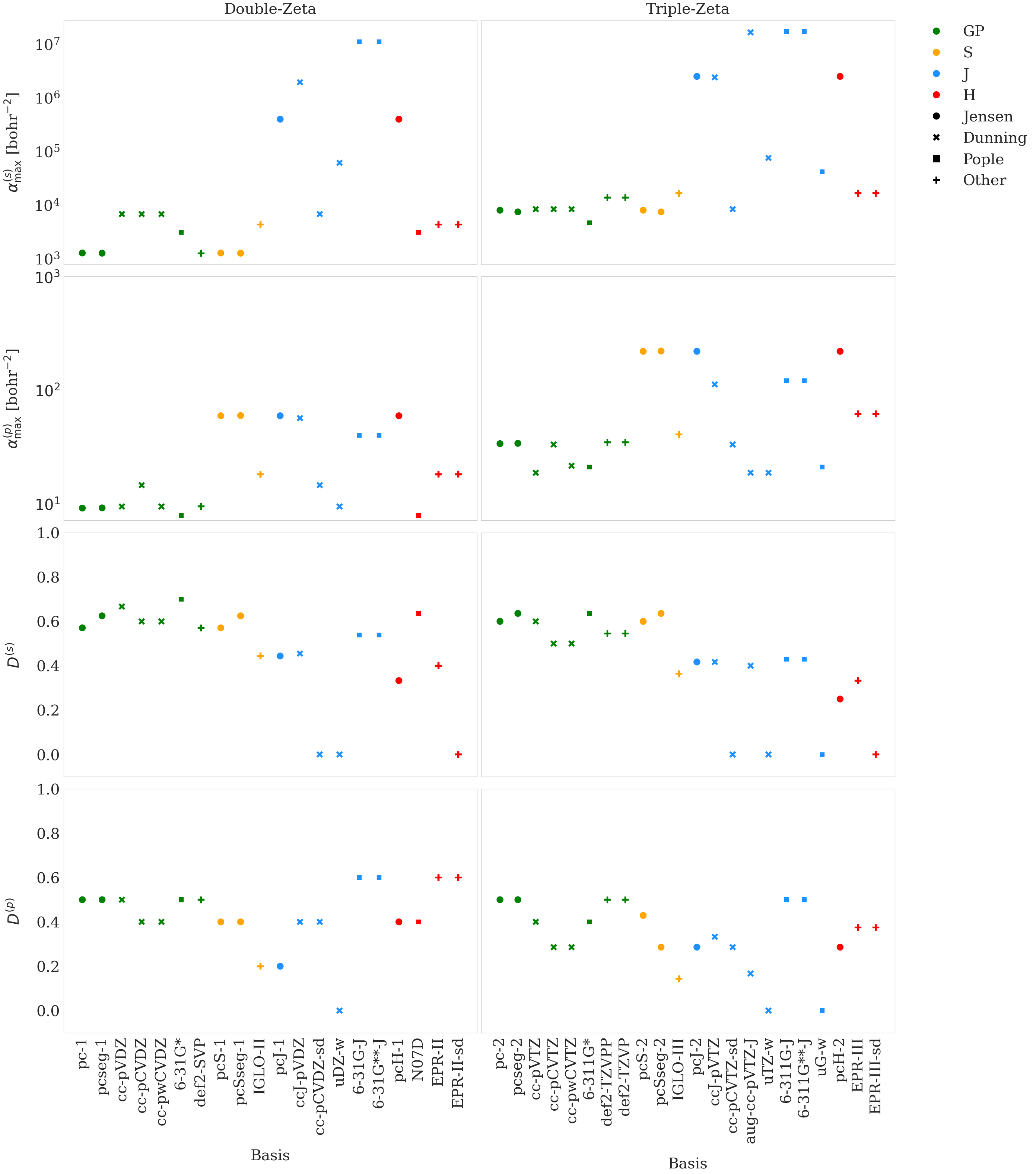}
    \caption{Largest exponent values, $\imax$, and degrees of contraction, $D^{(i)}$ for $s$- and $p$-functions of various double- and triple-zeta basis sets describing the carbon atom.}
    \label{fig:CarbonFigs}
\end{figure*}

\setlength{\tabcolsep}{5pt}
\renewcommand{\arraystretch}{1.2}
\begin{table*}[ht!] 
\centering
  \caption{Core-specialised basis sets of (double, triple)-zeta quality developed for particular core-dependent properties. Where applicable, the parent basis set from which they are derived is also provided. Brief notes on how the specialised set differs from its parent for the first row elements are also provided.}
  \begin{tabular}{@{}l c c c c @{}}
  \toprule
    &
    \multicolumn{1}{c}{Specialised Basis Set} &
    \multicolumn{1}{c}{Parent Basis Set} &
    \multicolumn{1}{c}{\hspace{0.4cm}Specialisation} &
    \\
    \midrule
    \mc{3}{l}{\textbf{Magnetic shielding constants}} \\
     & pcS-(1,2) \cite{Jensen2008} & pc-(1,2) & tight $p$ \\ 
     & pcSseg-(1,2) \cite{Jensen2015}  & pcseg-(1,2) & tight $p$, new $p$-contraction
     \\ 
     & IGLO-(II,III)\cite{Kutzelnigg1990, Huzinaga1971}   & None \\
     \vspace{-0.7em} \\
      \mc{3}{l}{\textbf{J coupling constants}} \\
   
     & pcJ-(1,2) \cite{Jensen2006, Aggelund2018}  & pc-(1,2) & tight $s$($\times$2), $p$ $d$, $f$, new $s$-, $p$-contraction 
     \\ 
     & ccJ-pV(D,T)Z \cite{Benedikt2008} & cc-pV(D,T)Z  & tight $s$($\times$2), $p$, $d$, new $s$-, $p$-contraction 
     \\
     & u(D,T)Z-w \footnote{The uDZ-w basis was generated for completeness in an analogous way to the uTZ-w basis set} & aug-cc-pV(D,T)Z & tight $s$($\times$w), complete decontraction      \\   
     & aug-cc-pVTZ-J \cite{Enevoldsen2001,Provasi2001,Provasi2010,Hedegard2011} & aug-cc-pVTZ & tight $s$($\times$4), remove diffuse $f$, new $s$-, $p$-contraction  \\
          \vspace{-0.7em} \\
      \mc{3}{l}{\textbf{J coupling constants (Fermi-Contact term only)}} \\
   
     & 6-31(1)G-J \cite{Kjr2011} & 6-31(1)G & tight $s$($\times$3), tight $p$, new $s$-, $p$-contraction \\
     & 6-31(1)G**-J \footnote{The 6-31(1)G**-J bases were generated from the 6-31(1)++G**-J bases \cite{Kjr2011} by removing diffuse functions} & 6-31(1)G** & tight $s$($\times$3), tight $p$, new $s$-, $p$-contraction \\
     & cc-pCV(D,T)Z-sd \cite{Peralta2004} & cc-pCV(D,T)Z & complete $s$-space decontraction \\
    & uG-w \cite{Deng2006} & 6-311+G(d,p) & tight $s$($\times$w), complete decontraction \\
     \vspace{-0.7em} \\
      \mc{3}{l}{\textbf{Hyperfine coupling constants}} \\
     & EPR-(II,III)\cite{Barone2002,Rega1996} & {Huzinaga-Dunning} & {tight $s$, additional polarisation functions, uncontracted inner-core} \\
     & EPR-(II,III)-sd\footnote{The EPR-III-sd basis was generated for completeness in analogous way to the EPR-II-sd basis set} \cite{Peralta2004} & {Huzinaga-Dunning} & Decontract EPR-(II,III) $s$-space \\
     & N07D \cite{Barone2008,Barone2009,Brancato2007Unraveling,Brancato2007Theoretical} & 6-31G & 6G core, polarisation/diffuse modifications element-dependent \\
     & pcH-(1,2)\cite{Jakobsen2019} & pc-(1,2) & tight $s$($\times$2), $p$ $d$, $f$, new $s$-, $p$-contraction  \\
    \bottomrule 
  \end{tabular}
  \vspace{-0.5em} \label{table:doubezetabasis}
\end{table*}

\subsection{Alternatives to Specialisation} \label{sec:Alternitives}

We note that there are alternative approaches that can be used for accurate calculation of core-dependent properties in a reasonably computationally-efficient manner. We briefly outline a few of these approaches below.

\begin{itemize}

    \item {\textbf{Locally-Dense Basis Sets:} The use of high-quality specialised basis sets over the whole molecule is not always a viable option computationally. The locally-dense basis sets approximation provides an alternative to this. This approximation involves using a large, high-quality basis set to describe only the target atom or group of atoms in a molecule for which the property of interest will be calculated. Then, lower-quality basis set(s) may be used to describe the remaining part(s) of the molecule \cite{Chesnut1989, DiLabio1998, Provasi2000, Theibich2021,Sanchez2005,O.Milhoj2014,Reid2014}. For example, Liang et. al have recently investigated calculating magnetic shielding constants by combining the locally-dense basis set approximation with composite methods \cite{Liang2023}. Depending on the desired accuracy/time constraints, different values of $n$ for the pcSseg-$n$ basis can be used to describe the target group, neighbouring groups and the remaining molecule.}

    \item {\textbf{Non-all-Gaussian Basis Sets:} It is possible to use basis sets which model correctly the electron behaviour at the nucleus. Mixed ramp-Gaussian basis sets \cite{McKemmish2014, McKemmish2015a}, basis sets which consist of a polynomial function (the `ramp') until 1 Bohr radius and Gaussian functions thereafter, are an example of such basis sets. The ramp function is beneficial as it satisfies the Kato cusp condition \cite{Kato1957}, with limited test calculations showing their effectiveness at describing the electron density at nuclear positions \cite{mckemmish2015accurate} whilst maintaining good performance for valence-dependent properties \cite{McKemmish2015a,Cox2021}. Crucially, initial testing of integral evaluation performance shows that ramp-Gaussian basis sets can be calculated in times competitive with all-Gaussian basis sets \cite{McKemmish2015a}. }

    \item {\textbf{Scaling Factors:} The use of scaling factors may be used as an alternative (or in addition) to the use of specialised basis sets and may serve as a route to remove some of the difficulties associated with specialising basis sets. In general, scaling factors involve the comparison of experimental and theoretical values to define a factor with which other theoretical values can be scaled by to reduce errors. Scaling factors have been calculated and applied to shielding constants \cite{LodewykComputational2012, Pierens2014, Iron2017, MerrillSolvent2020}, and J coupling constants \cite{Suardiaz2008,SanFabian2013}. }
\end{itemize}

\subsection{Evaluating the Description of Functional Spaces}

Specialisation is no easy task; as alluded to in Sec \ref{subsec:Specialised}, different operators will probe different functional spaces \cite{Jensen2006}. Typically, one considers all of the terms contributing towards a particular property. However, one may wish to focus on improving only the functional space(s) which contribute the largest errors.

In order to get a sense of how well a functional space is described, the largest Gaussian exponents in a basis set $\imax$ (for $i=$ $s$, $p$, $d$ etc...) can be examined. The higher the value of $\imax$, the better the quality of the description of functional space $i$ near the nucleus.

Furthermore, it is useful to mathematically quantify the extent of contraction in a basis set. Consider an uncontracted basis set with $N^{(i)}_{\ur{u}}$ primitive functions of type $i$. Upon contraction, this reduces to the number of contracted (basis) functions, $N^{(i)}_{\ur{c}}$. The contraction ratio $C^{(i)}$ can then be defined as 
\begin{align}
    C^{(i)} = \frac{N^{(i)}_{\ur{c}}}{N^{(i)}_{\ur{u}}} \ , \label{eq:ConRatio}
\end{align}
from which we define the degree of contraction $D^{(i)}$ as
\begin{align}
    D^{(i)} = 1 - C^{(i)} \ . \label{eq:DoC}
\end{align}

For example, the cc-pVQZ basis is contracted as $12s6p3d2f1g \to 5s4p3d2f1g$, meaning that $D^{(s)} \approx 0.583$ and $D^{(p)} \approx 0.333$ and so indicating that the $s$-functions are contracted to a greater extent than the $p$-functions. The value of $D^{(i)}$ takes into account the fact that larger basis sets have more functions. That is, a large basis sets needs to contract more primitive functions to get the same degree of contraction as a smaller basis set, and so reduce the quality of the description of the functional space by a comparable amount.

There is no single approach to designing basis sets for particular properties. To demonstrate this, the largest $s$- and $p$-exponents ($\alpha_{\ur{max}}^{(i)}$) and degrees of contraction ($\iD$) for the specialised basis sets given in Table \ref{table:doubezetabasis} have been collected. The results for hydrogen and carbon (taken to be representative of the first row elements) are presented in Figs. \ref{fig:HydrogenFigs} and \ref{fig:CarbonFigs}, respectively. The basis sets are collected in terms of the property they are specialised towards: magnetic shieldings (S), J couplings (J) and hyperfine couplings (H). A number of general-purpose basis sets (GP) are also included for comparison. The $\iD$ value for each basis has been calculated using \rref{eq:DoC}. {We note that $\imax$ and $\iD$ do not encapsulate basis set design in its entirety. However, these are the parameters which are altered during specialisation and, as will be discussed, highlight well the differences between general-purpose basis sets and the various core-specialised basis sets studied.}

Figures \ref{fig:HydrogenFigs} and \ref{fig:CarbonFigs} make immediately clear that general-purpose basis sets and core-specialised basis sets are designed in very different ways. Furthermore, basis sets specialised towards different core-properties are also designed in different ways. This is on account of the fact that different properties probe different parts of the molecular wave function.  This can further be shown by considering the description of different functional spaces using the completeness profile \cite{Delano1995Coompleteness, Manninen2006Systematic}. 

The completeness profile $\mathcal{Y}\lrr{\alpha}$ measures the `completeness' of a functional space by measuring its overlap with a test function. In our case, the test functions are taken to be an $s$-GTO for the $s$-space and a $p$-GTO for the $p$-space. The completeness profile is valued between zero and unity and is measured as a function of the logarithm of the exponent $\alpha$ of the test function. Therefore, large $\log_{10}\lrr{\alpha}$ values can be considered to be representative of the core region, whilst lower values are representative of the valence region. 

\begin{figure}[thbp!]
    \centering
    \hspace*{-0.5cm}\includegraphics[scale = 0.2]{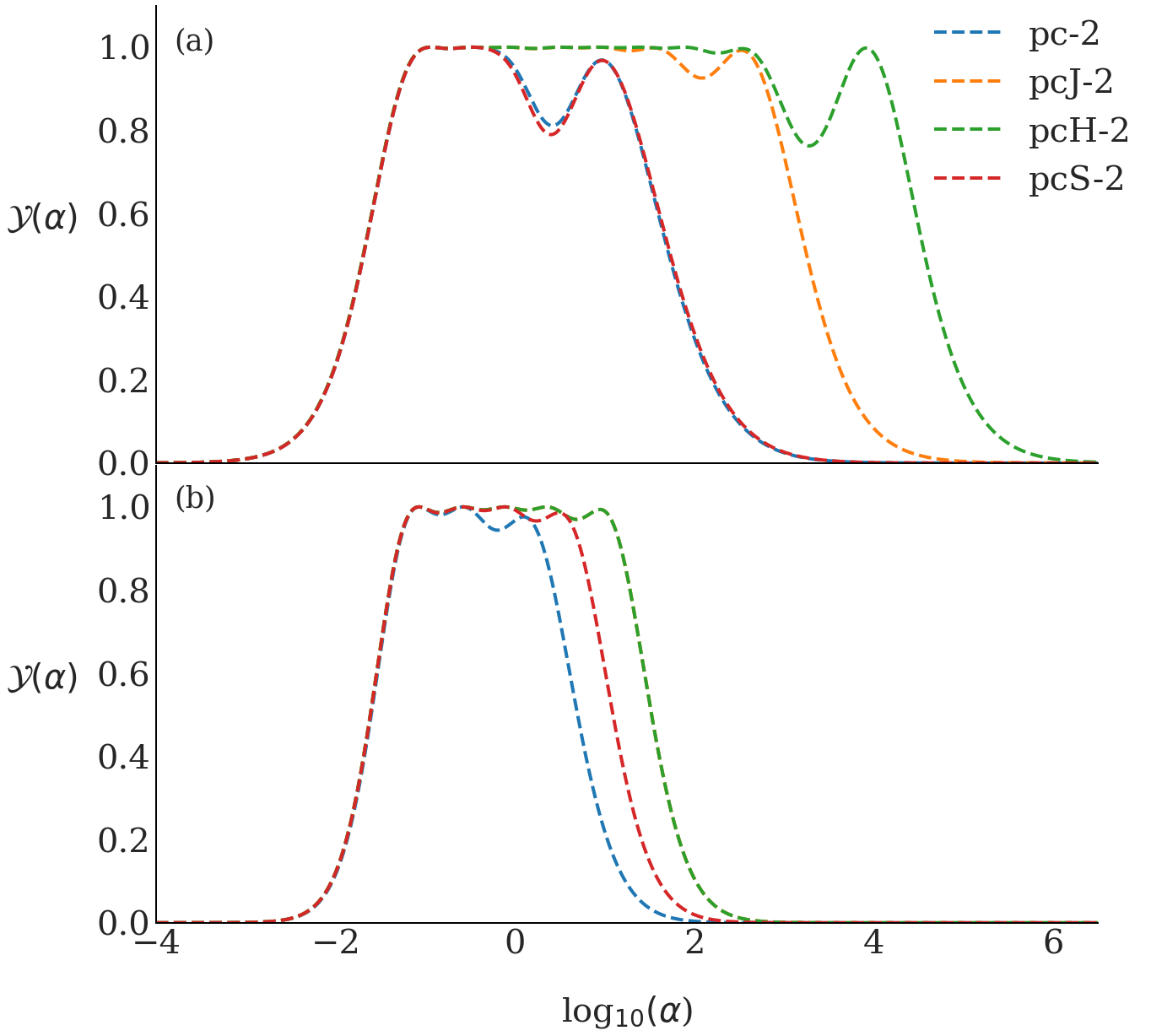}
    \caption{Completeness profiles for the (a) $s$- and (b) $p$-spaces of various Jensen-style basis sets representing the carbon atom. Note that the $p$-spaces of the pcJ-2 (orange) and pcH-2 (green) bases overlap.}
    \label{fig:CompletnessProfiles}
\end{figure}

To show that different properties probe different functional spaces, Fig. \ref{fig:CompletnessProfiles} shows the completeness profiles for the $s$- and $p$-spaces of four Jensen-style basis sets. Specifically, we study the pcJ-2, pcH-2, pcS-2 and pc-2 basis sets. The first three sets are specialised towards the calculation of J coupling constants, hyperfine coupling constants and magnetic shielding constants, respectively, whilst the final set is the parent general-purpose basis set. 
Despite all the specialised Jensen sets investigated stemming from the pc-2 basis, Fig. \ref{fig:CompletnessProfiles} shows that completeness of the functional spaces described by these bases differ on account of how they have been specialised. This is because specialisation aims to improve the description of the specific part of the functional space(s) relevant to the core property being probed.

We note the identical nature of the regions further from the nucleus (low $\log_{10}\lrr{\alpha}$) of all of the basis sets in Fig. \ref{fig:CompletnessProfiles}, which is that inherited from the parent pc-2 basis; the specialised bases only require enhanced descriptions of the region near and at the nuclear position, meaning the valence region can remain unaltered.

With reference to Fig. \ref{fig:CompletnessProfiles}, the basis with the best described $s$-space near the nucleus is the pcH-2 basis. This is consistent with Fig. \ref{fig:CarbonFigs} which shows pcH-2 to have the lowest $\sD$ values of the Jensen basis sets. The pc-2 and pcS-2 bases have the least-well described $s$-spaces near the nucleus, which is supported by these bases having amongst the lowest $\smax$ and highest $\sD$ values.  The pcH-2 and pcJ-2 bases have well-described $p$-spaces, given their low $\pD$ values. The pcS-2 basis also does well in its description of the $p$-space, given that this basis set is specialised by the addition of a tight $p$-function.

Referring back to Fig. \ref{fig:HydrogenFigs}, we see that the only commonality between the specialised basis sets is that the $\pD$ values for hydrogen are zero for almost all basis sets. This, however, is likely on account of the fact that smaller basis sets are typically used for hydrogen, and so is unlikely to be due to specialisation reasons. The fact that the general-purpose basis sets also have this feature supports this claim.

\section{Benchmark Calculation Methodology}
\label{sec:Methodology}

\subsection{Basis Set Selection}

We will investigate specialised sets derived from the polarisation-consistent (Jensen) \cite{Jensen2001, Jensen2002a, Jensen2003, Jensen2004}, correlation-consistent (Dunning and co.) \cite{Dunning1989,Kendall1992, Woon1993, Woon1994, Woon1995, Wilson1996,Wilson1999, VanMourik2000, Dunning2001,  Prascher2011} and Pople \cite{Ditchfield1970SelfOrbitals, HehreSelf1970, Ditchfield1971, Hehre1972SelfBoron, Hehre1972SelfElements, Hehre1972, Hariharan1973,Krishnan1980SelfFunctions, Franci1982SelfElements,Frisch1984SelfSets} basis set families, limiting our investigations to sets of double- and triple-zeta quality. The specialised basis sets selected are shown in Table \ref{table:doubezetabasis}.  For each property, a number of general-purpose basis sets are also used in order to contextualise our results. 
 
 Thorough benchmarking of basis sets for general-purpose chemistry is not readily available, though we refer the reader to the preliminary benchmarking from Pitman \etal{} \cite{PitmanBenchmarking}. Of particular note for our purposes here is the conclusion that the 6-311G family of basis sets is not of true triple valence-zeta quality and is probably more accurately described as double core-zeta and double valence-zeta quality. Though this wasn't the design intention, this increased flexibility in the core has the potential to allow 6-311G** to perform more strongly for core properties than is typical for most general-purpose basis sets (although this is balanced by far worse valence chemistry predictions).

 We note that specialised basis sets have also been developed for other core-dependent properties, such as core-electron binding energies \cite{Hanson-Heine2018, Fouda2018, Ambroise2019,  Foerster2020, Ambroise2021} which can be probed experimentally using X-ray photoelectron spectroscopy (XPS). However, the basis sets for this property in particular are dependent on the method by which the basis sets are designed, and so we exclude these sets from our discussion.
 
 For our studies, we also limit ourselves to hydrogen, oxygen, carbon and nitrogen- (HONC-)containing molecules. This eliminates the study of some basis sets such as the acv$X$Z-J sets, designed specifically for the calculation of J coupling constants for selenium, tellurium \cite{rusakov2019a} and tin \cite{Rusakov2021}, and basis sets specialised towards iron-57 in M\"ossbauer spectroscopic calculations \cite{Neese2002,Kurian2010, Casassa2016}.

\subsection{Level of Theory Selection} 

Our study focuses on double and triple-zeta basis sets, so the level of theory should  have similar errors and thus we consider pure Hartree-Fock and various density functional theories. Note that double-hybrid DFT and coupled-cluster wave function methods should use basis set extrapolation techniques based on larger basis sets to avoid the basis set errors being significantly larger than the level of theory error.

Looking in the literature at the performance of different levels of theory, we note:

\begin{itemize}
    \item {\textbf{J coupling:} The exchange-correlation functional has been found to play an important role in the calculation of J coupling constants \cite{Malkin1994Calculation,FabianImprovements2014}, with the Fermi-contact term (see Sec. \ref{Sec:JTheory}) term being particularly sensitive to the choice of exchange-correlation functional \cite{PatchkovskiiCuring2001,KealGIAO2004}. The B3LYP functional has often been found to perform well \cite{Sychrovsky2000Nuclear, Helgaker2000Analytical, PerttuSpin2002, Suardiaz2008, SanFabian2013},  as well as functionals from the B97 family \cite{KupkaBasis2012}. }

    \item {\textbf{Hyperfine coupling:} Like with J coupling constants, the hyperfine coupling constant can also be sensitive to the choice of functional \cite{BatraCalculations1996} likely due also to the presence of the Fermi-contact term (see Sec. \ref{Sec:HTheory}). The difficulties of using functionals to calculate hyperfine coupling constants for radical species is well known \cite{MunzarovaCritical1999,ImprotaInterplay2004, HermosillaDFT2011, WitwickiDensity2018, WitwickiHow2020}. The common B3LYP functional has often been recommended for hyperfine coupling constants in radicals of various size \cite{BatraCalculations1996, ZakrassovPredictions2002, NguyenEfficent1997, NguyenCalculation1997}, whilst the suitability of the MO6 suit of functionals for \ce{^{14}N} hyperfine constants \cite{GromovPerformance2019} and the B2PLYP functional for some transition metals \cite{KossmanPerformance2007} has also been noted. }

    \item {\textbf{Shielding constants:} When concerned with organic molecules and molecules which are not strongly correlated, Hartree-Fock and `pure' DFT approaches can provide better results than common density functionals when calculating magnetic shielding constants \cite{MagyarfalviAssesment2003, HieringerDensity2004, KupkaConvergence2010, LodewykComputational2012}. An overestimation in the paramagnetic terms (see Sec. \ref{Sec:STheory}) due to underestimating the HOMO-LUMO gap can cause an underestimation in shielding constants \cite{WilsonDensity1999,MagyarfalviAssesment2003,HieringerDensity2004,TealeBenchmarking2013,KupkaConvergence2010}, though functionals have been designed to correct for this underestimation. For example, the KT1 and KT2 functionals produce accurate shielding constants \cite{KealExchange2003,KealGIAO2004,TealeBenchmarking2013}, and the improved KT3 functional which performs comparably to common GGAs for thermochemistry calculations \cite{KealSemiempirical2004}.}
\end{itemize}

Calculations in our study for each property were performed using pure Hartree-Fock theory and density functional theory using four functionals: the Hartree-Fock-Slater (HFS) exchange only functional; the B3LYP hybrid functional; the meta-GGA M06-2X hybrid functional; and the range-separated $\omega$B97X-D3 hybrid functional. The use of multiple levels of theory ensures our conclusions focus on basis set performance and are fairly robust to method choice. 

We do not report individual method results in the main paper as our selection was modest. Interested readers can refer to the Supplementary Material.

\subsection{Molecular Benchmark Data set} \label{sec:MoleBench}

\begin{table*}[htpb!] 
\centering
  \caption{Molecules considered in the isotropic J coupling constants benchmarking. Molecules in bold were used to assess basis set timings.}
  \begin{tabular}{@{}l c c c c c c  c c c  @{}}
    \toprule
     & $\ur{CH_4}$ & $\ur{C_2H_2}$ & $\ur{CH_2NH}$ & $\ur{{NO_3}^-}$& $\ur{C_2H_4}$  & $\ur{C_2H_6}$ & \textbf{Peroxyacetyl Nitrate}  &  \textbf{Pyridine} & \textbf{Imidazole} 
     \\
      & $\ur{{H_2NO_2}^-}$ & $\ur{CO}$ & $\ur{{H_2NO_2}^+}$ & $\ur{HCN}$& $\ur{CH_2O}$  & Glycine & \textbf{Cytosine} & \textbf{Urea} & \textbf{Alanine} 
     \\
       & $\ur{O_2}$& $\ur{H_2O}$ & $\ur{N_2}$ & $\ur{NH_3}$ & $\ur{CH_3OH}$  & \textbf{Uracil} & \textbf{Thymine}
     \\
     \bottomrule
     \label{table:JCoupMolecules}
  \end{tabular}
\centering
  \caption{Molecules considered in the isotropic hyperfine coupling constants benchmarking. Molecules in bold were used to assessbasis set timings.}
  \begin{tabular}{@{}l c c c c c c @{}}
\toprule
    & CH   & $\ur{{CO_2}^-}$ &  $\ur{{O_2}^-}$& $\boldsymbol{\ur{(CH3)_2CCN}}$ & \textbf{Dimethylaminomethyl Radical}\\
     & $\ur{CH_2}$ & $\ur{CN}$ & $\ur{HO}$  & $\boldsymbol{\ur{(CH_3)_2CCONH_2}}$ & \textbf{3-Amino-2-Propenyl Radical}\\
     
     & $\ur{CH_3}$  & $\ur{CHO}$  & $ \ur{HO_2} $ & \textbf{Diformylaminyl Radical}  & \textbf{Methylmethoxyamino Radical}\\
     & $\ur{C_2H_3}$  & $\ur{CHO_2}$& $\ur{NO_2}$ &  Methyl Amino Radical  &  \textbf{4-Imidazoline-2-yl Radical} \\
     & $\ur{C_2H_5}$ & $\ur{H_2NO}$& $ \ur{H_2NO_2}$&  Aminomethyl Radical &  \textbf{Glycine Radical} \\
     
\bottomrule
    \label{table:HypeFineMolecules}
  \end{tabular}
\centering
  \caption{Molecules considered in the isotropic magnetic shielding constants benchmarking. Molecules in bold were used to assessbasis set timings.}
  \begin{tabular}{@{}l c c c c c c c c c@{}}
  
\toprule
      
      & $\ur{{CO_2}^-}$& $\ur{CH_4}$  & $\ur{C_2H_3}$ &  $\ur{C_2H_5}$  & $\ur{CH_2O}$   &  Cytosine  & \textbf{3-Amino-2-Propenyl Radical}\\ 
      & $\ur{N_2}$ & $\ur{NH_3}$& $\ur{CN}$ & $\ur{HCN}$ & $\ur{CH_3CN}$ & \textbf{Alanine} & Aminomethyl Radical \\
      & $\ur{O_2}$ & $\ur{{O_2}^-}$ & $\ur{CH_3}$ & $\ur{CH_2NH}$  &$\ur{CH_3CH_2OH}$ & Glycine & Methylmethoxyamino Radical \\
      & $\ur{HO}$ &$\ur{C_2H_2}$ & $\ur{C_2H_4}$ &$\ur{C_2H_6}$ & Urea & \textbf{Thymine}&  Dimethylaminomethyl Radical \\ 
        & $\ur{CO_2}$  & $\ur{CHO_2}$  & $\ur{CH_2O}$ & $\ur{CH_3OH}$ & $ \boldsymbol{\ur{(CH3)_2CCN} }$ & \textbf{Uracil} & \textbf{Peroxyacetyl Nitrate}  \\
        
& $ \ur{H_2NO_2} $ & $\ur{{NO_3}^-}$ & $\ur{\lrr{CH_3}_3N}$ & $\ur{CH_3NO_2}$&  $\boldsymbol{\ur{(CH_3)_2CCONH_2}}$  & Pyridine& Diformylaminyl Radical \\
      &   $\ur{{H_2NO_2}^+}$ & $\ur{NO_2}$  & $\ur{HCOOH}$ & $\ur{CH_3COOH}$  & & Imidazole& \textbf{4-Imidazoline-2-yl Radical}  \\
& $\ur{H_2O}$ & $\ur{H_2NO}$  & $\ur{CH_2}$&  & & Glycine Radical & Methyl Amino Radical  \\
     \bottomrule
     \label{table:ChenShieldMolecules}
  \end{tabular}
\end{table*}

Our goal was to select a modest set of small- to medium-sized HONC-containing molecules relevant for each property. We were looking for typical expected performance and thus excluded molecules from consideration that were found to be particularly challenging for all basis sets, specifically $\ur{{NO_2}^+}$, CO, CHO, $\ur{HO_2}$ and CH. These difficult cases can be considered further in the future, alongside more extensive benchmark sets such as NS372 for chemical shieldings \cite{SchattenbergExtended2021}.

The molecules selected are similar to those used in the original optimisation of the property-specialised basis sets.  Molecules were carefully chosen, however, to ensure that there would not be a bias towards one particular basis set's training data set. In total, 47 molecules were studied for isotropic shielding constants, whilst 25 and 23 were studied for isotropic hyperfine couplings and isotropic J couplings, respectively. The set of molecules used for isotropic shielding constants, isotropic J coupling constants and isotropic hyperfine coupling constants can be found in Tables \ref{table:JCoupMolecules}, \ref{table:HypeFineMolecules} and \ref{table:ChenShieldMolecules}, respectively.  Geometries for all molecules were optimised at the $\omega$B97M-V/pc-2 level.
 
\subsection{Computational Details} \label{sec:CompDetails}

 All calculations were performed using ORCA 5.0.0 \cite{Neese2012, Neese2022}.  Basis sets that were not included in the ORCA library were either collected from Basis Set Exchange \cite{feller1996a,Schuchardt2007,Pritchard2019} or inferred from the literature. 
 
 For the calculation of magnetic shielding constants, gauge-including atomic orbitals (GIAO) were used in all calculations. Furthermore, the calculation of J coupling constants was done for all pairs of atoms separated by no more than 5 \AA. 

 Since our aim is to provide practical user recommendations on good basis set selection for specialised properties, some model chemistry results were excluded for some properties. Specifically, the uG-w basis failed to provide results in a number of instances, and so was removed from the data sets. Furthermore, the uDZ-w basis gave extremely unrealistic results in J coupling calculations, and so is not included in the analysis of isotropic J coupling constants. The Hartree-Fock results for isotropic J coupling constants were also not included for similar reasons. 

 \subsection{Data Analysis} \label{sec:DataAnalysis}
 Results in this benchmark are not compared with experimental results, as this would require the careful control of other variables such as solvation and vibration effects. Instead, benchmark values were calculated using the fully uncontracted specialised analogues of the pc-4 basis -- that is, the pcJ-4 basis for isotropic J couplings, the pcH-4 basis for isotropic hyperfine couplings and the pcSseg-4 basis for isotropic nuclear magnetic shieldings. The fully uncontracted versions of these sets containing diffuse functions were not used for calculating benchmark values as SCF convergence issues were often encountered, a well documented issue \cite{Chandrasekhar_Efficient_1981, Papajak_Perspectives_2011, Bauza_Is_2013, PitmanBenchmarking}.  

 Using the reference benchmark values calculated, we calculated the statistical distribution of errors, mean absolute deviations (MADs) and median values for the relevant set of molecules for each basis set. All results are reported in the Supplementary Information.
 
 In the main paper, we focus on considering median results rather than MADs as is more standard. The reasoning for this is that there is a significant number of outliers and we want to focus on providing users with details of expected performance.  

In order to show the consistency of a basis with respect to the different levels of theory, the range of values for the median errors are plotted for each basis set. This range is defined using the functional with the greatest median error as the upper-limit, and the functional with the lowest median error as the lower limit. We refer to this as the median error range (MER). 

The MER is used in an attempt to assess the consistency a basis set achieves with respect to the different levels of theory. However, it is appreciated that a single large error associated with a particular level of theory would lead to a misleadingly-large MER, especially if all other levels of theory returned lower errors. For this reason, the median value of the median errors associated with each level of theory is also shown for each basis, which we refer to as the median of the MER. Figures containing all median errors for each functional explicitly can be found in the Supporting Information.

Results for each property are separated based on the magnitude of the median errors; results for isotropic shielding constants are separated with respect to hydrogen and non-hydrogen shielding constants; results for J coupling constants are separated with respect to H:H, X:H and X:X couplings (where X= O, N, C); and results for hyperfine coupling constants are separated element-wise.

\subsection{Timings}

In determining the most appropriate computational method for a given situation, the calculation time is a critical factor. Calculation times are very dependent on molecule size and the specific computational chemistry package (in particular whether the program takes advantage of shared exponents in generally contracted basis sets or $sp$-shells). Nevertheless, it is useful to consider relative timings for a subset of typical molecules. These molecules range in size from 7 to 15 atoms (4 to 9 non-hydrogen atoms) and are shown in bold in Tables \ref{table:JCoupMolecules} - \ref{table:ChenShieldMolecules}.

Timings for each basis set are measured relative to the timings for the popular 6-31G** basis for each property and are averaged over the subset of molecules discussed. All timings relate to calculations using the $\omega$B97X-D3 functional. Total times take into account not only the calculation of the property of interest, but also SCF iterations and the evaluation of Gaussian integrals. Furthermore, timings for the calculation of J coupling constants and hyperfine coupling constants also include the calculation magnetic shielding tensors. 

All timings reported here should be considered indicative only.

\section{Isotropic J coupling Constants} \label{Sec:IsoJCoupling}
\subsection{Theoretical Background} \label{Sec:JTheory}

The J coupling interaction, also referred to as spin-spin coupling, is a relatively weak interaction which arises due to interacting nuclear spins in a molecule. Based on the alignment of spins, the interaction can cause nuclear spins to be shielded or deshielded from the applied magnetic field. Note that this shielding is of a different form to that discussed in Sec. \ref{sec:IsoShieldingConstants}. The resulting impact of the J coupling interaction on spectra is that spectral peaks can be `split' into further sub-peaks. 

The J coupling interaction provides information about molecular structure in both the solution- and solid-states, though in the case of the latter the peaks usually remain unresolved.  We shall for this reason focus on J coupling from the perspective of the solution state, where due to motion averaging only the isotropic component is observed. 

The J coupling tensor $\bd{J}$ is a $3 \times 3$ tensor containing the information regarding J coupling interactions, where the isotropic component is found as $\frac{1}{3}\Tr{\textbf{J}}$. Overall, the J coupling tensor takes the form 
\begin{align}
    \textbf{J} = 
    \bd{J}^{\ur{DSO}} + \bd{J}^{\ur{PSO}} + \bd{J}^{\ur{FC}} + \bd{J}^{\ur{SD}} \ .\label{eq:JTensor}
\end{align}
where $\bd{J}^{\ur{DSO}}$ is the diamagnetic spin-orbit (DSO) term, $\bd{J}^{\ur{PSO}}$ is the paramagnetic spin-orbit (PSO) term, $\bd{J}^{\ur{FC}}$ is the Fermi-contact (FC) term and $\bd{J}^{\ur{SD}}$ is the spin-dipole (SD) term, all defined below \cite{Ramsey1953,Pyykko2000}. 
With the exception of the DSO term, these terms are calculated pertubatively, meaning that \textbf{J} can be written as
\begin{align*}
    \bd{J} = \bd{J}^{\ur{DSO}} + \bd{J}^{\ur{P}} \ , \nr
\end{align*}
where 
\begin{align}
    \textbf{J}^\ur{P} \propto
    \sum_{M} \sum_{n>0} \frac{\matrixel{\Psi_0}{\hat{H}^M_A}{\Psi_n}\matrixel{\Psi_n}{\hat{H}^M_B}{\Psi_0}}{E_0 - E_n} \ . \label{eq:Response} 
\end{align}
Here, $M$ refers to either SD, PSO or FC, and so $\hat{H}^M_{A/B}$ refers to the operator corresponding to the relevant contribution with respect to the $A^{\ur{
th}} / B^{\ur{th}}$ nucleus. The $\Psi_0$ state is the ground state with energy $E_0$, whilst $\Psi_n$ represents the $n^{\ur{th}}$ excited state with energy $E_n$. The operators corresponding to each of the contributions take the form
\begin{align}
    \hat{H}^{\ur{PSO}} &\propto \sum_i \frac{\bd{r}_{iA}\cross \nabla_i}{r_{iA}^3} \ , \label{eq:PSO} \\ 
    \hat{H}^{\ur{SD}} &\propto \sum_i \frac{r_{iA}^2\bd{s}_i - 3\lrr{\bd{s}_i \cdot \bd{r}_{iA}}\bd{r}_{iA}}{r_{iA}^5} \ , \label{eq:SD} \\
    \hat{H}^{\ur{FC}} &\propto \sum_{i}\bd{s}_i\delta\lrr{\br_{iA}} \ . \label{eq:FC}
\end{align}
The DSO term is calculated using the DSO operator,
\begin{align}
    \bd{J}^{\ur{DSO}} = \matrixel{\Psi_0}{\hat{H}^{\ur{DSO}}}{\Psi_0} \ , \label{eq:JDSO}
\end{align}
with
\begin{align*}
    \hat{H}^{\ur{DSO}} &\propto  
    \sum_{i} \frac{\lrr{\br_{iA}^T\cdot \br_{iB}}\mathbbm{1}_3 - \lrr{
    \br_{iA}\otimes\br_{iB}^T}}{r_{iA}^3r_{iB}^3} \  , \nr \label{eq:DSO}
\end{align*}
where we have that $\br_{iA}$ represents the separation between electron $i$ and nucleus $A$, $\br_{iB}$ represents the separation between electron $i$ and nucleus $B$, $\nabla_i$ is the nabla operator for electron $i$, $\bd{s}_i$ is the spin angular momentum operator for electron $i$, $\delta\lrr{x}$ is the Dirac-delta distribution, $\mathbbm{1}_3$ is the $3 \times 3 $ unit matrix and the superscript $T$ denotes the vector transpose. 

The Dirac-delta distribution present in the FC term (also present in the calculation of the hyperfine coupling tensor tensor, Sec. \ref{Sec:HTheory}) probes the wave function at the nuclear position. The FC term is often the dominating term in such calculation \cite{Jensen2006, Aggelund2018}, making an accurate description of core electrons even more vital. The denominators of the DSO, PSO and SD term (all effective $r^{-3}$ terms) indicate that the electron description in the vicinity of the nucleus is of importance. Furthermore, the perturbative manner in which many of the terms are evaluated suggests that the region further from the nucleus will also be important, i.e. that a poor valence basis set description could lead to poor predictions. 

\subsection{Basis Set Demands and Design} \label{Sec:JDemands}

Our analysis of basis set design in Figs. \ref{fig:HydrogenFigs} and \ref{fig:CarbonFigs} show that J-specialised basis sets typically have larger maximum $s$-exponents ($\smax$ values) and smaller $s$-degrees of contraction ($\sD$ values) than their general-purpose counterparts. 
Interestingly Figs. \ref{fig:HydrogenFigs} and \ref{fig:CarbonFigs} show that basis sets specialised for J coupling constants actually follow two distinct designs principles: 
\begin{enumerate}[label={(\arabic*)}]
    \item Intermediate $\sD$ values,  high $\smax$ and typically high $\pmax$ (pcJ-$n$, ccJ-pV$n$Z, aug-cc-pVZT-J); these typically focus on all terms in the J coupling tensor. \label{cat1}
    
    \item Zero $\sD$ values (i.e. no contraction of $s$-primitives), smaller $\smax$  and typically low $\pmax$ (u(D,T)Z-w, uG-w, cc-pCV(D,T)Z-sd, 6-31(1)G(**)-J); these typically focus on only the FC term in the J coupling tensor. \label{cat2}
\end{enumerate} 

Similar to the behaviour of the electron density at nuclear positions \cite{Mastalerz2010}, it has been noted frequently that the FC term is sensitive to the presence of tight {and loosely-contracted} $s$-functions. As an explanation for the elements of the basis set design, Jensen \cite{Jensen2006} also noted that the PSO and DSO terms are sensitive to presence of tight $p$-functions while the SD term required tight $p$-, $d$- and $f$- functions. These dependencies arise due to the difference in the operators involved in the calculation of J coupling constants. As such, some basis sets will make an attempt to focus solely on the dominating FC term (which may be useful when larger systems are of interest \cite{Kjr2011, Colasuonno2020}), whilst others try to account for all the terms present in the J coupling tensor. 

We note from Figs. \ref{fig:HydrogenFigs} and \ref{fig:CarbonFigs} that the sets with the lowest $\smax$ and $\sD$ values (category \ref{cat2} basis sets above) also tend to have the lowest $\pmax$ values too, indicating that these sets are focused on the description of the core region and so are primarily concerned with the FC term. Conversely, those with the larger $\smax$ and intermediate $\sD$ values (category \ref{cat1} basis sets above) tend to also have larger $\pmax$ values, indicating that these basis sets are focused on the description of the core and core-valence regions and so are concerned with all terms in the J coupling tensor.

\subsection{Basis Set Performance} \label{Sec:JPerformance}

\begin{figure}[thbp!]
    \centering
    \includegraphics[scale = 0.19]{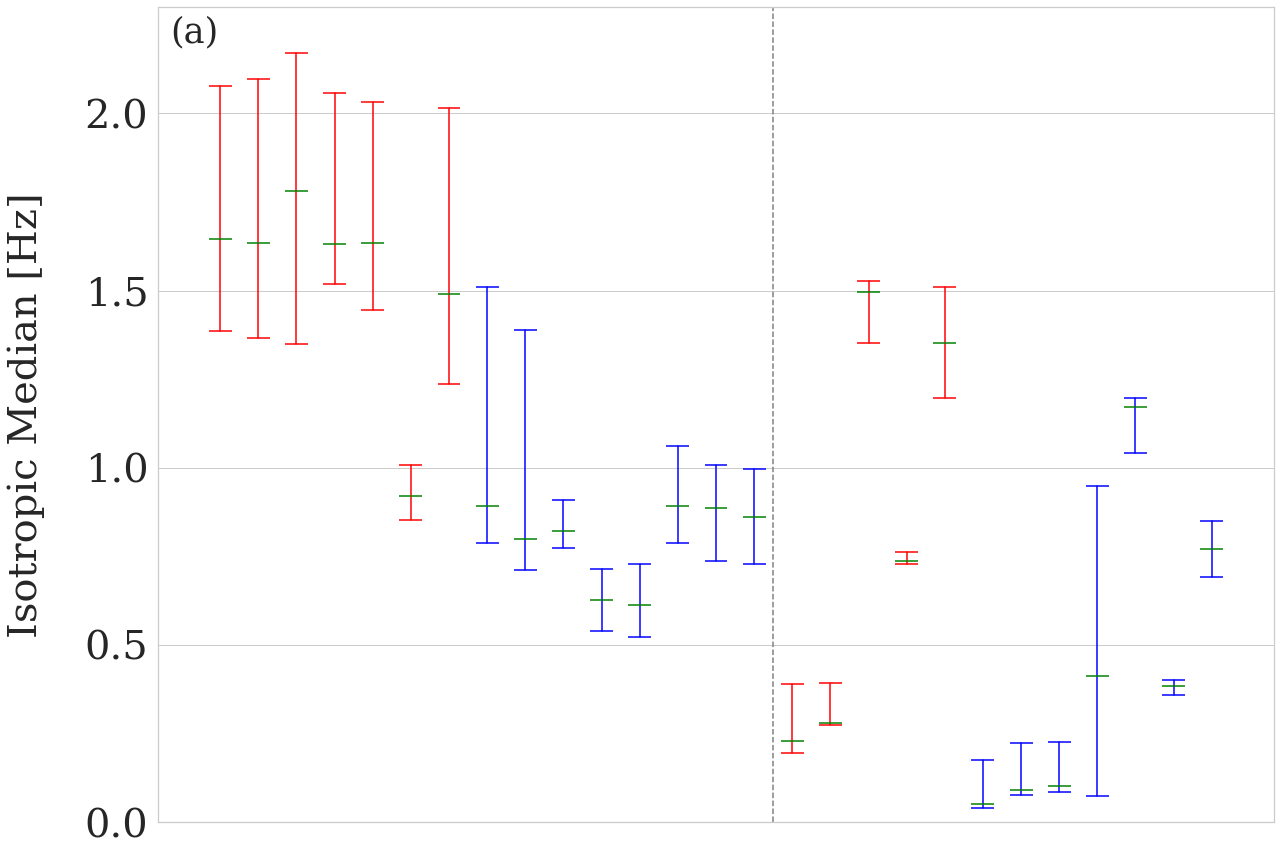}
        \vspace{0.5em}
        
    \includegraphics[scale = 0.19]{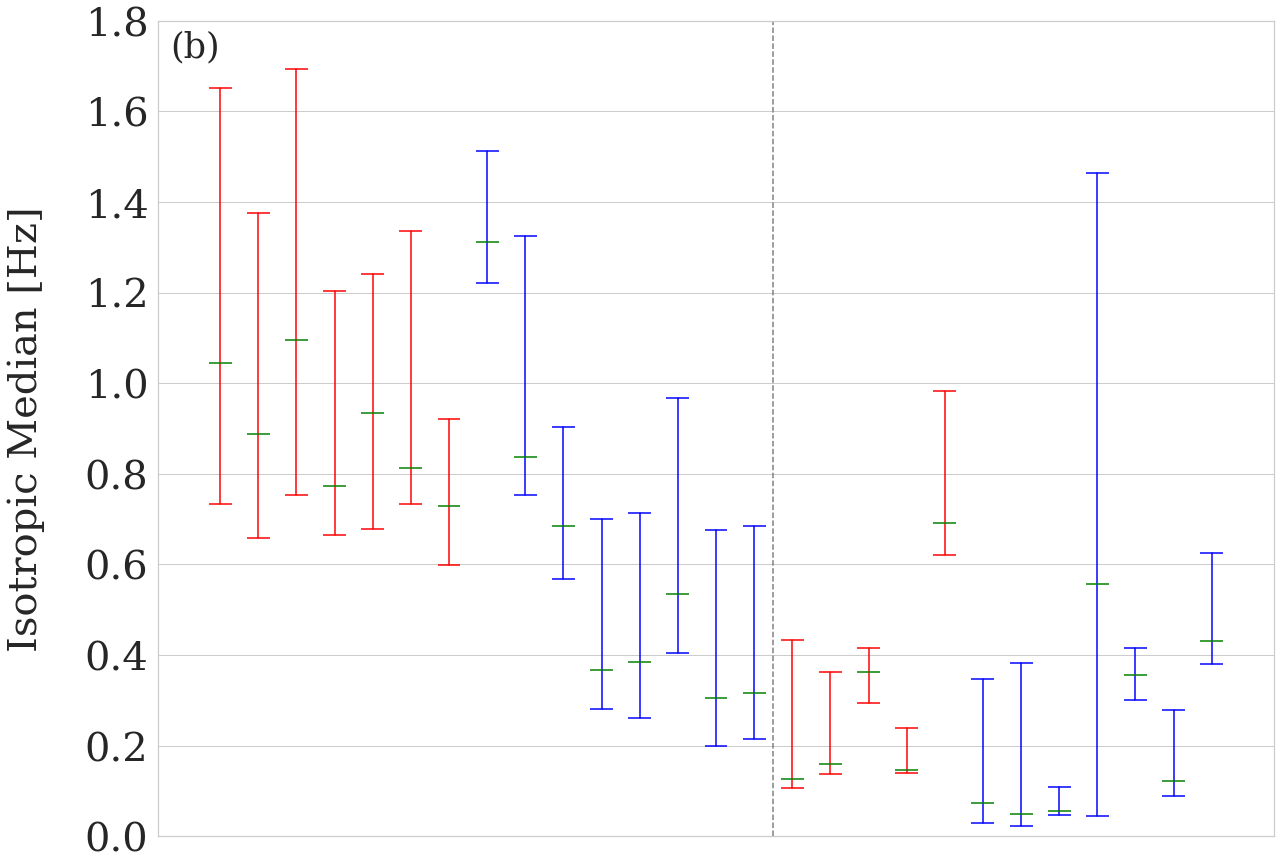}
    \vspace{0.5em}
    
    \includegraphics[scale = 0.19]{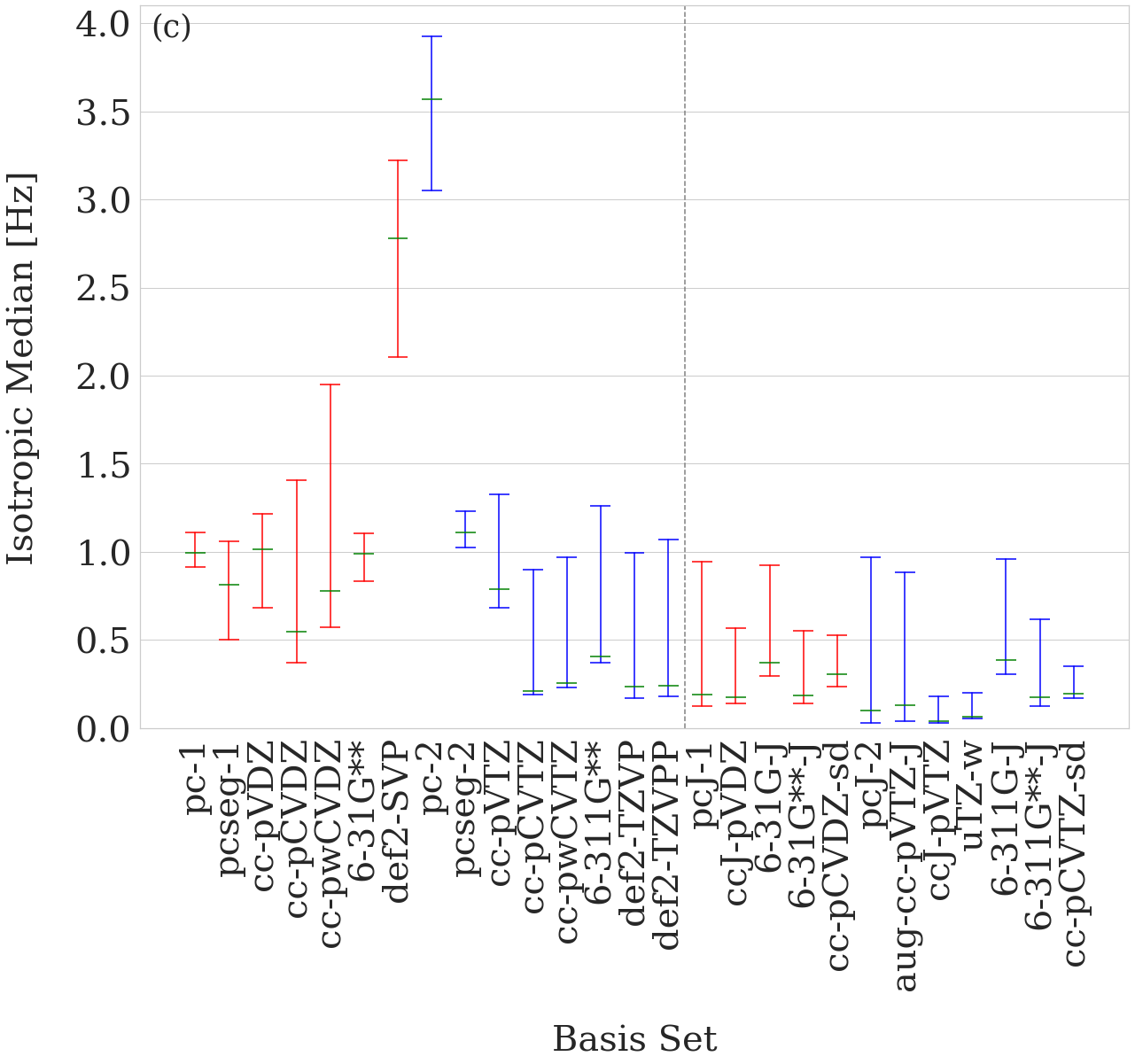}
    \caption{The median error ranges (MERs) for isotropic J coupling constants for (a) H:H (b) X:H and (c) X:X couplings. Basis sets to the left of the horizontal grey line are general-purpose basis sets, whilst those to the right are J-specialised basis sets. Ranges in red present double-zeta basis sets, whilst those in blue represent triple-zeta basis sets. For each basis, the median values of the MERs are given as the green horizontal lines.}
    \label{fig:JCouplingMedianErrors}
\end{figure}

The median error ranges (MERs) and median MER values for isotropic J coupling constants are shown in terms of H:H, X:H and X:X couplings in Figs. \ref{fig:JCouplingMedianErrors} (a), (b) and (c), respectively \footnote{Despite the cc-pV$X$Z, cc-pCV$X$Z and cc-pwCV$X$Z sets being identical for hydrogen, differences are observed between these sets for the H:H couplings. This suggests an influence from other atoms in the molecule, though this analysis is outwith the remit of this paper.}.

It is immediately obvious that many J-specialised basis sets significantly outperform the general-purpose basis sets, with median MER reducing typically by a factor of 2-5 depending on the coupling. In fact, double-zeta specialised basis sets have performance generally similar to triple-zeta general-purpose basis sets.

We can correlate the basis set performance with basis set design by comparing Fig. \ref{fig:JCouplingMedianErrors} to Figs. \ref{fig:HydrogenFigs} and \ref{fig:CarbonFigs}. It is clear that the best-performing sets across all couplings are those with large $\smax$ values and modest $\sD$ values (category \ref{cat1} above).  The importance of a well-described core-valence and valence regions can be appreciated here, as for the non-hydrogen elements the triple-zeta pcJ-2, ccJ-pVTZ and aug-cc-pVTZ-J basis sets all have polarisation functions up to and including $l = 3$, $f$-functions ($l=2$, $d$-functions for the double-zeta equivalents). The significant improvement in performance between 6-311G-J and 6-311G**-J further shows that polarisation functions are important for high accuracy, which help with the description of the core-valence and valence regions. The influence of basis set design on basis set performance is described in more detail in the Supporting Information. 

\begin{table*}[htpb!]
\centering
\caption{Timings for the basis sets used in the calculation of isotropic J coupling constants. Timings are reported relative to those of the 6-31G** basis set (underlined). The J-specialised basis sets are shown in bold. Contracted basis descriptions, as well as the number of basis functions per atom ($N_{\ur{A}}$), for the first row elements are also shown. These timings should be considered as indicative only.}
\begin{tabular}{@{}l c c c c c @{}}
\toprule &
Double-Zeta  & Triple-Zeta &  Rel. Timing & Contracted Basis & $N_{\ur{A}}$ \\
\midrule
        &  \textbf{6-31G-J} &                   &               \textbf{0.8} & $\boldsymbol{6s2p}$ & \textbf{12} \\
        & def2-SVP &                   &               0.9 & ${3s2p1d}$ & 14\\
        &  \underline{6-31G**} &                   &               \underline{1.0} & $\underline{{3s2p1d}}$ & $\underline{14}$\\
        &  pcseg-1 &                   &               1.0 & ${3s2p1d}$ & 14 \\
        &     pc-1 &                   &               1.1 & $3s2p1d$ & 14 \\
        &          &          \textbf{6-311G-J} &               \textbf{1.1} & $\boldsymbol{8s3p}$ & \textbf{17}\\
        &  cc-pVDZ &                   &               1.2 & $3s2p1d$ & 14 \\
        &          &          6-311G** &               1.4 & $4s3p1d$ & 18 \\
        & cc-pCVDZ &                   &               1.4 & $4s3p1d$ & 18 \\
    &    \textbf{6-31G**-J} &                   &               \textbf{1.4} & $\boldsymbol{6s2p1d}$ & \textbf{17} \\
     &   cc-pwCVDZ &                   &               1.5 & $4s3p1d$ & 18 \\
      &            &        \textbf{6-311G**-J} &               \textbf{1.7} & $\boldsymbol{8s3p1d}$ & \textbf{22} \\
       &     \textbf{pcJ-1} &                   &               \textbf{1.8} & $\boldsymbol{5s4p2d}$ & \textbf{27} \\
        &          &         def2-TZVP &               2.4 & $5s3p2d1f$ & 31 \\
         &         &        def2-TZVPP &               3.0 & $5s3p2d1f$ & 31 \\
         &         &           pcseg-2 &               3.4 & $4s3p2d1f$ & 30 \\
         &         &           cc-pVTZ &               3.5 & $4s3p2d1f$ & 30 \\
         &         &              pc-2 &               3.5 & $4s3p2d1f$ & 30\\
      & \textbf{cc-pCVDZ-sd} &                   &               \textbf{4.0} & $\boldsymbol{10s3p1d}$ & \textbf{24} \\
    &     \textbf{ccJ-pVDZ} &                   &               \textbf{4.4} & $\boldsymbol{6s3p2d}$ & \textbf{25} \\
     &             &          cc-pCVTZ &               4.9 & $6s5p3d1f$ & 43 \\
      &            &         cc-pwCVTZ &               5.2 & $6s5p3d1f$ & 43\\
       &           &             \textbf{pcJ-2} &               \textbf{6.6} & $\boldsymbol{7s5p3d2f}$ & \textbf{51}\\
        &          &     \textbf{aug-cc-pVTZ-J} &               \textbf{7.0} & $\boldsymbol{9s5p3d1f}$ & \textbf{46} \\
         &         &          \textbf{ccJ-pVTZ} &              \textbf{13.1} & $\boldsymbol{7s4p3d1f}$ & \textbf{41} \\
          &        &       \textbf{cc-pCVTZ-sd} &            \textbf{14.3} & $\boldsymbol{12s5p3d1f}$ & \textbf{49} \\
           &       &             \textbf{uTZ-w} &              \textbf{34.0} & $\boldsymbol{13s6p3d2f}$ & \textbf{59} \\
\bottomrule\label{Table:JTimings}
\end{tabular}
\end{table*}

 Table \ref{Table:JTimings} shows the average timings involved in the calculation of the full J coupling tensors, as well as the magnetic shielding tensor, for each basis set relative to the 6-31G** basis. It is important to note that using J-specialised or core-specialised basis sets always increases the calculation time, sometimes quite dramatically; Table \ref{Table:JTimings} shows that some specialised double-zeta basis sets are slower than general-purpose triple-zeta basis sets for our system size. 

Considering both performance and timing, the triple-zeta bases that fall into category \ref{cat1} above are computationally cheaper than those in catagory \ref{cat2}, as is reported in Table \ref{Table:JTimings}, suggesting that this design is more efficient on account of the somewhat-contracted $s$-spaces. Specifically, the J-specialised Jensen basis sets, pcJ-1 (double-zeta, faster) and pcJ-2 (triple-zeta, more accurate), are clearly the best choice of basis sets for J coupling calculations and should be used in the future.

\section{Isotropic Hyperfine Coupling Constants} \label{Sec:IsoHyperfine}
\subsection{Theoretical Background} \label{Sec:HTheory}
In many ways, one can view hyperfine coupling as the equivalent to J coupling for electron paramagnetic resonance (EPR) spectroscopy; hyperfine coupling results from the interaction between the spins of unpaired electrons and nuclei. This interaction splits fine structure energy levels (determined by electron spin) into further sub-levels.  

This splitting effect is described by the hyperfine coupling tensor, $\bd{A}$, which in its full form is written in terms of three components, the (isotropic) Fermi-contact (FC) term $\bA ^{\ur{FC}}$, the (anisotropic) spin-dipole (SD) term $\bA ^{\ur{SD}}$ and the second order spin-orbit (SO) coupling contribution $\bA^{\ur{SO}}$. The hyperfine coupling tensor can be described overall using a nonrelativistic pertubative framework as \cite{Sauer2011, Kaupp2004, Woodgate1983}
\begin{align}
    \bA = \bA ^{\ur{FC}} + \bA^{\ur{SD}} + \bA^{\ur{SO}} \ . \label{eq:HyperfineTensor}
\end{align}
The SD and FC terms are calculated using the following operators, 
\begin{align}
        \hat{H}^{\ur{SD}} &\propto \sum_i \lrr{\frac{r_{iA}^2\bd{s}_i - 3\lrr{\bd{s}_i \cdot \bd{r}_{iA}}\bd{r}_{iA}}{r_{iA}^5}}\cdot \bd{I}_A \ , \label{eq:ASD} \\
    \hat{H}^{\ur{FC}} &\propto \sum_{i}\bd{s}_i \cdot \bd{I}_A\delta\lrr{\br_{iA}} \ , \label{eq:AFC}
\end{align}
where $\bd{I}_A$ is the spin operator for nucleus $A$. We note that Eqs. (\ref{eq:ASD}) and (\ref{eq:AFC}) take very similar forms to those given in Section \ref{Sec:IsoJCoupling}. Hence, we would expect that the basis set demands for H-specialised basis sets are not dissimilar to those for J-specialised basis sets.

The SO term is usually negligible when there are no heavy elements present; we do not calculate it for our benchmarking here. Furthermore, the anisotropic SD term averages to zero in solution. Thus, we will focus here on the isotropic component, the FC term relating to \rref{eq:AFC}, which tends to be the dominant term.

\subsection{Basis Set Demands and Design} \label{Sec:HDemands}

The $s$-basis set functional space is most important to describe the Fermi contact term accurately; higher angular momentum functions only contribute from other nuclei and this will be very small. Thus hyperfine-specialised (H-specialised) basis sets typically have low $\sD$ and high $\smax$, like J-specialised basis sets, as shown in Figures \ref{fig:HydrogenFigs} and \ref{fig:CarbonFigs}. 

Basis sets are designed to also reproduce the anisotropic SD term (though we don't provide benchmarking results for this property here). This is achieved through the addition of tight $p$-, $d$- and $f$-basis functions \cite{Jakobsen2019}, which typically increases $\pmax$ and decreases $\pD$.  


\begin{figure*}[htbp!]
    \centering
    \includegraphics[scale = 0.2]{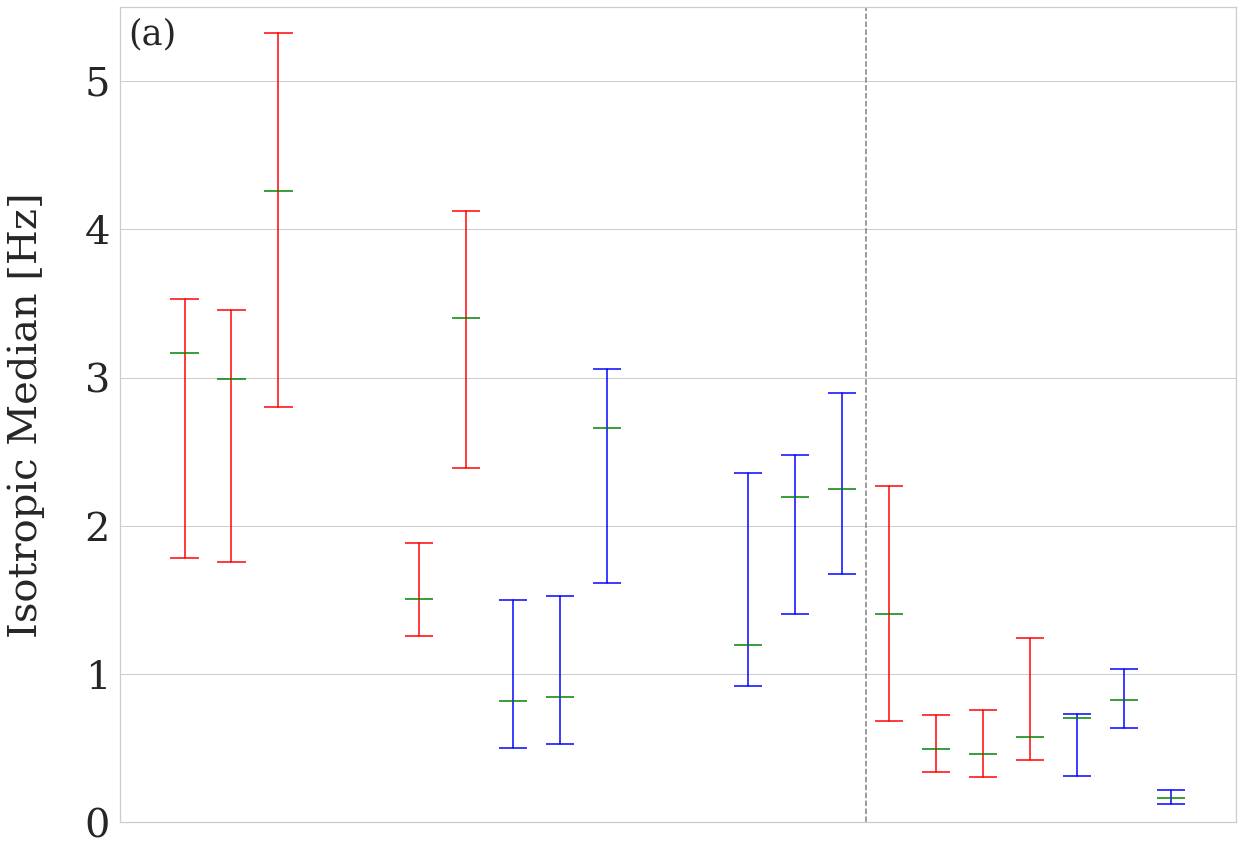}
    \includegraphics[scale = 0.2]{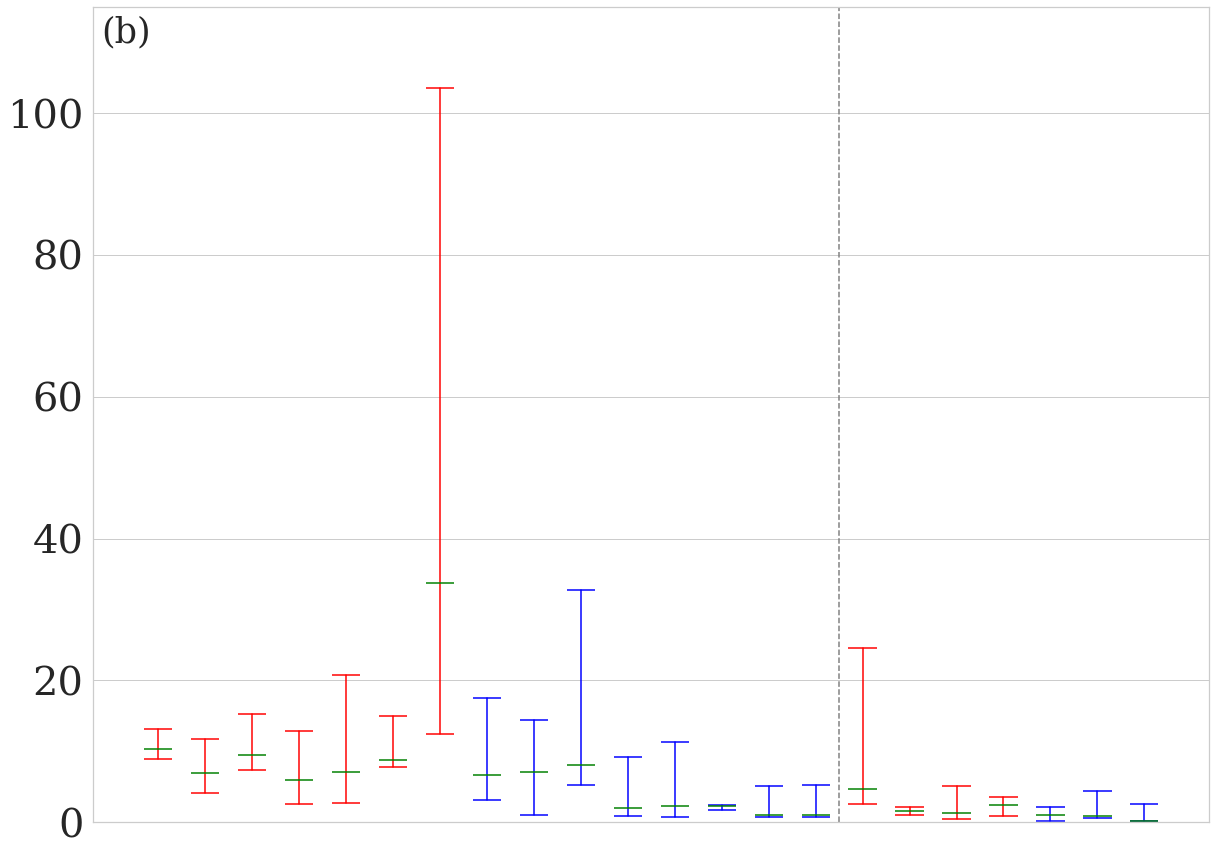}
    \hspace*{-0.14cm}\includegraphics[scale = 0.2]{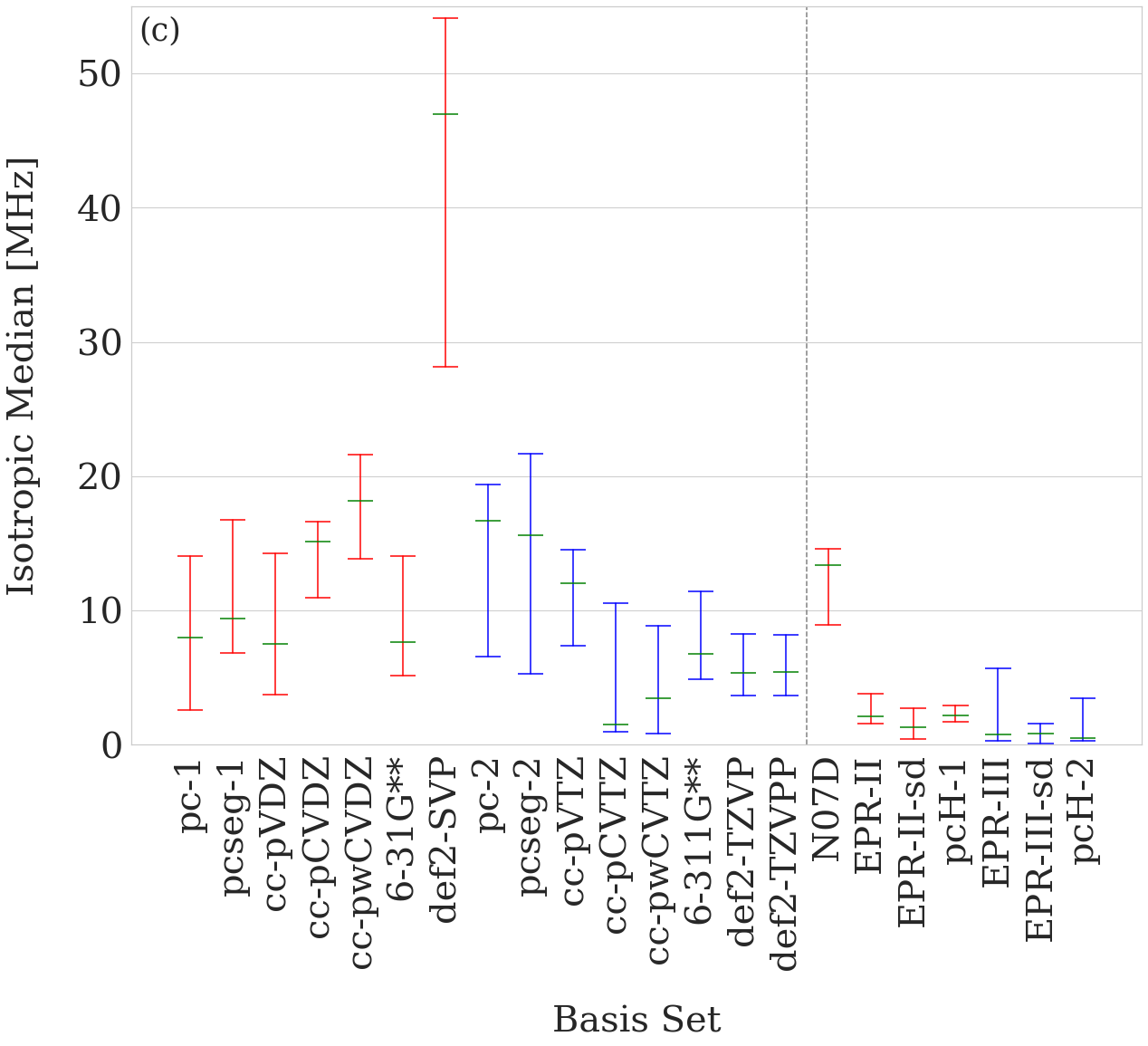}
    \hspace*{0.14cm}\includegraphics[scale = 0.2]{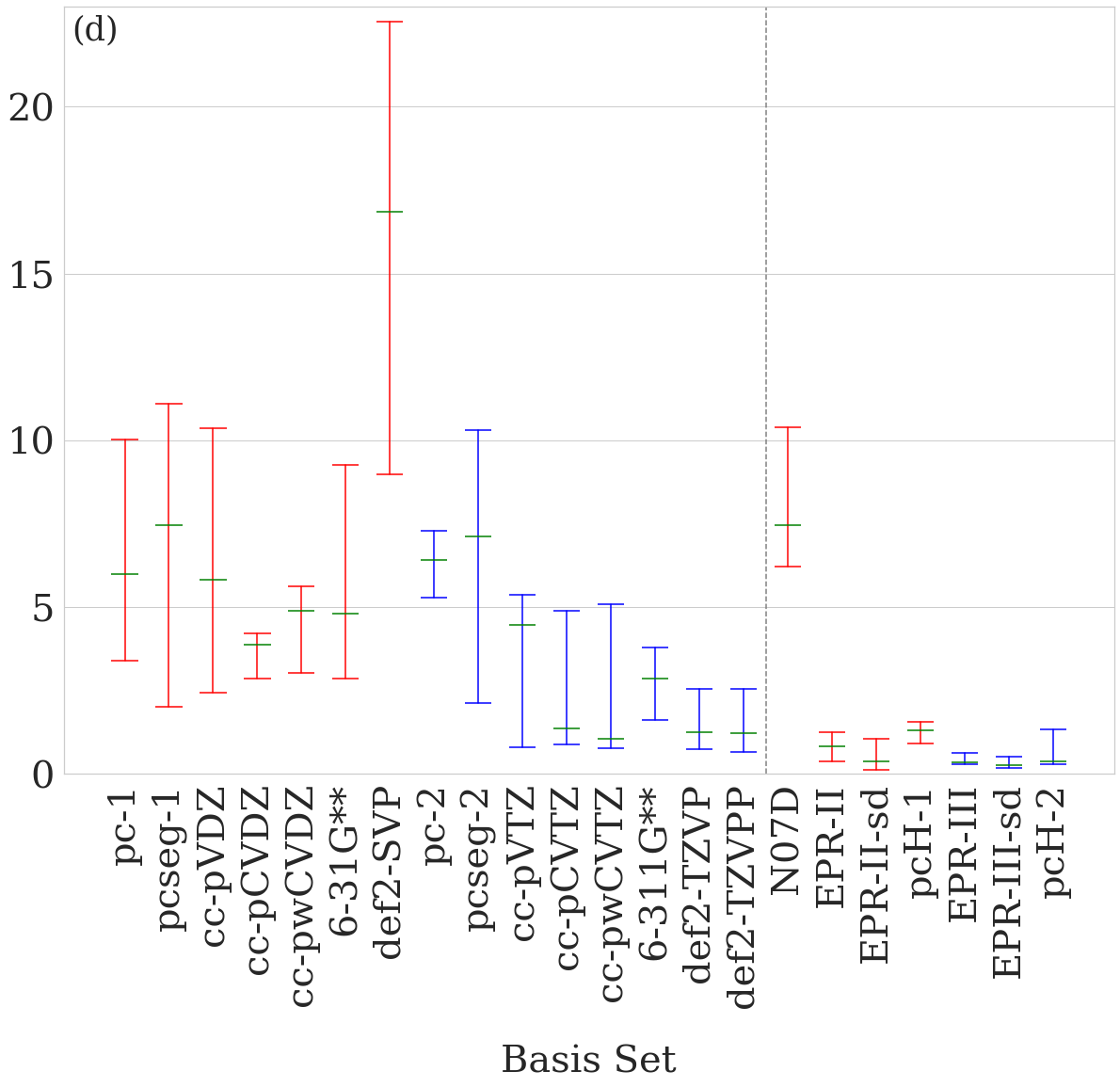}
    \caption{The median error ranges (MERs) for isotropic hyperfine coupling constants for (a) hydrogen (b) carbon (c) oxygen and (d) nitrogen. Basis sets to the left of the horizontal grey line are general-purpose basis sets, whilst those to the right are H-specialised basis sets. Ranges in red present double-zeta basis sets, whilst those in blue represent triple-zeta basis sets. For each basis, the median values of the MERs are given as the green horizontal lines.}
    \label{fig:HyperfineMedianErrors}
\end{figure*}

\subsection{Basis Set Performance} \label{Sec:HPerformance}

The median error ranges (MERs) for isotropic hyperfine coupling constants for hydrogen, oxygen, carbon and nitrogen are shown in Figs. \ref{fig:HyperfineMedianErrors} (a) - (d), respectively. 

Like for J coupling, the H-specialised basis sets outperform general-purpose basis sets for all elements; indeed, we even see significant improvement in most cases between general-purpose triple-zeta and specialised double-zeta basis sets, consistent with the FC term describing the isotropic tensor. 

Strong-performing basis sets for the calculation of hyperfine coupling constants have high $\smax$ and low $\sD$ values. Basis sets with lower $\smax$ and higher $\sD$ values tend to under-perform, as is the case with the general-purpose and specialised N07D bases. It is interesting to note that the 6-31G** and 6-311G** bases perform relatively well in all circumstances. This is despite a heavier contraction of its $s$-space than other general-purpose sets. In the case of 6-311G**, the fact that the basis is actually double zeta in its core \cite{Cox2020} is most likely responsible for its enhanced performance. In general however, high $\smax$ values also appear to be important and can alleviate some of the need to decontract the $s$-space, though low $\sD$ values are still required. More information regarding the design of H-specialised basis sets can be found in the Supporting Information. 

\begin{table*}[htbp!]
\centering
\caption{Timings for the basis sets used in the calculation of isotropic hyperfine coupling constants. Timings are reported relative to those of the 6-31G** basis set (underlined). The H-specialised basis sets are shown in bold. Contraction schemes, as well as the number of basis functions per atom ($N_{\ur{A}}$), for the first row elements are also shown. These timings should be considered as indicative only.}
\begin{tabular}{@{}l c c c c c @{}}
\toprule &
Double-Zeta  \hspace{0.2cm} & Triple-Zeta  \hspace{0.2cm} &  Rel. Timing & Contracted Basis & $N_{\ur{A}}$ \\
\midrule
          &\underline{6-31G**} &                   &             \underline{1.0} & $\underline{3s2p1d}$ & \underline{14}\\
          &pcseg-1 &                   &             1.1 & $3s2p1d$ & 14 \\
         &def2-SVP &                   &             1.1 & $3s2p1d$ & 14\\
          &        &          6-311G** &             1.1 & $4s3p1d$ & 18 \\
         &    pc-1 &                   &             1.2 & $ 3s2p1d$ & 14\\
         & cc-pCVDZ &                   &             1.2 & $4s3p1d$ & 18 \\
        & cc-pwCVDZ &                   &             1.3 & $4s3p1d$ & 18 \\
        &  cc-pVDZ &                   &             1.3 & $3s2p1d$ & 14 \\
        &     \textbf{N07D} &                   &             \textbf{1.6} & $\boldsymbol{4s3p1d}$ & \textbf{18} \\
        &   \textbf{EPR-II} &                   &             \textbf{1.6} & $\boldsymbol{6s2p1d}$ & \textbf{17} \\
        &   \textbf{ pcH-1} &                   &             \textbf{1.8} & $\boldsymbol{6s3p2d}$ & \textbf{25} \\
        & \textbf{EPR-II-sd} &                   &             \textbf{2.4} & $\boldsymbol{10s2p1d}$ & \textbf{21} \\
         &         &           pcseg-2 &             2.6 & $4s3p2d1f$ & 30 \\
         &         &          cc-pCVTZ &             2.6 & $6s5p3d1f$ & 43 \\
         &         &         cc-pwCVTZ &             2.8 & $6s5p3d1f$ & 43  \\
         &         &              pc-2 &             3.1 & $4s3p2d1f$ & 30 \\
         &         &         def2-TZVP &             3.5 & $5s3p2d1f$ & 31 \\
         &         &           \textbf{EPR-III} &             \textbf{4.2} & $\boldsymbol{8s5p2d1f}$ & \textbf{40}\\
         &         &        def2-TZVPP &             4.3 & $5s3p2d1f$ & 31 \\
         &         &           cc-pVTZ &             5.3 & $4s3p2d1f$ & 30 \\
         &         &        \textbf{EPR-III-sd} &             \textbf{7.2} & $\boldsymbol{12s5p2d1f}$ & \textbf{44} \\
         &         &             \textbf{pcH-2} &            \textbf{12.3} & $\boldsymbol{9s5p3d2f}$ & \textbf{53}\\
\bottomrule
\label{Table:HyperfineTimings}
\end{tabular}
\end{table*}

The relative times taken to complete the calculation of isotropic hyperfine coupling constants, in addition to the evaluation of the magnetic shielding tensor, are given in Table \ref{Table:HyperfineTimings}. Specialising the basis set for hyperfine couplings adds significantly to the computational time due to the lower degree of contraction and additional primitives. Of particular note is the increase in computational cost resulting from complete decontraction of the $s$-space. This can be clearly seen by considering the additional time required when using the EPR-III-sd set (7.2 relative timing) compared to the EPR-III set (4.2 relative timing). Interestingly, this change leads to no noticeable improvement for the non-hydrogen elements indicating that the $s$-space for the EPR-III basis is well spanned and so complete decontraction has little (if any) impact.

Considering both time and performance, the EPR-III set is preferred for high-accuracy results. Though the pcH-2 basis set has slightly better performance, particularly for hydrogen, the calculation time increases significantly to almost three-times that of the EPR-III basis. 

For faster calculations using specialised double-zeta basis sets, the pcH-1 and EPR-II double-zeta sets also perform relatively well for hydrogen and the first row elements, and are only marginally more expensive computationally than the 6-31G** basis.

\section{Isotropic Magnetic Shielding Constants} \label{sec:IsoShieldingConstants}
\subsection{Theoretical Background} \label{Sec:STheory}

The resonance frequencies of nuclei in molecules are highly dependent on the electronic environment in the molecule. In NMR studies, the magnetic field experienced by a nucleus is not only due to that of the applied magnetic field, but also due to local magnetic fields resulting from the motion of electrons in the molecule. The difference between the magnetic field at a nuclear position and the applied magnetic field is incorporated into the magnetic shielding tensor, $\boldsymbol{\sigma}$. 

In addition to the PSO operator encountered in the J coupling tensor, \rref{eq:PSO}, the calculation of $\boldsymbol{\sigma}$ employs the diamagnetic shielding (DS) and orbital Zeeman (OZ) operators which take the form
\begin{align*}
    \hat{H}^{\ur{DS}} &\propto  
    \sum_{i} \frac{\lrr{\br_{iG}^T\cdot \br_{iA}}\mathbbm{1}_3 - \lrr{
    \br_{iA}\otimes\br_{iG}^T}}{r_{iA}^3} \  , \nr \label{eq:DiamagneticShielding} \\ 
    \hat{H}^{\ur{OZ}} &\propto \bd{r}_{iG} \cross \nabla_i \ , \nr  \label{eq:OrbitalZeeman}
\end{align*}
respectively, where $\bd{r}_{iG}$ is the separation between electron $i$ and the gauge origin $G$. The shielding tensor is then calculated as
\begin{align}
    \boldsymbol{\sigma} = \matrixel{\Psi_0}{\hat{H}^{\ur{DS}}}{\Psi_0} - 2\sum_{n\ne 0} \frac{ \matrixel{\Psi_0}{\hat{H}^{\ur{PSO}}}{\Psi_{n}} \matrixel{\Psi_n}{\hat{H}^{\ur{OZ}}}{\Psi_0} }{E_0 - E_n} \ . \label{eq:ShieldingConstant}
\end{align}
Similar to the J coupling tensor, the diamagnetic contribution can be calculated as a simple expectation value, whilst the other terms are evaluated pertubatively. 

As with the other properties discussed, only the isotropic component is observed in solution, which is evaluated as $\frac{1}{3}\Tr{\boldsymbol{\sigma}}$.

\subsection{Basis Set Demands and Design} \label{Sec:ShieldDemands}
\begin{figure*}[htbp!]
    \centering
    \includegraphics[scale=0.2]{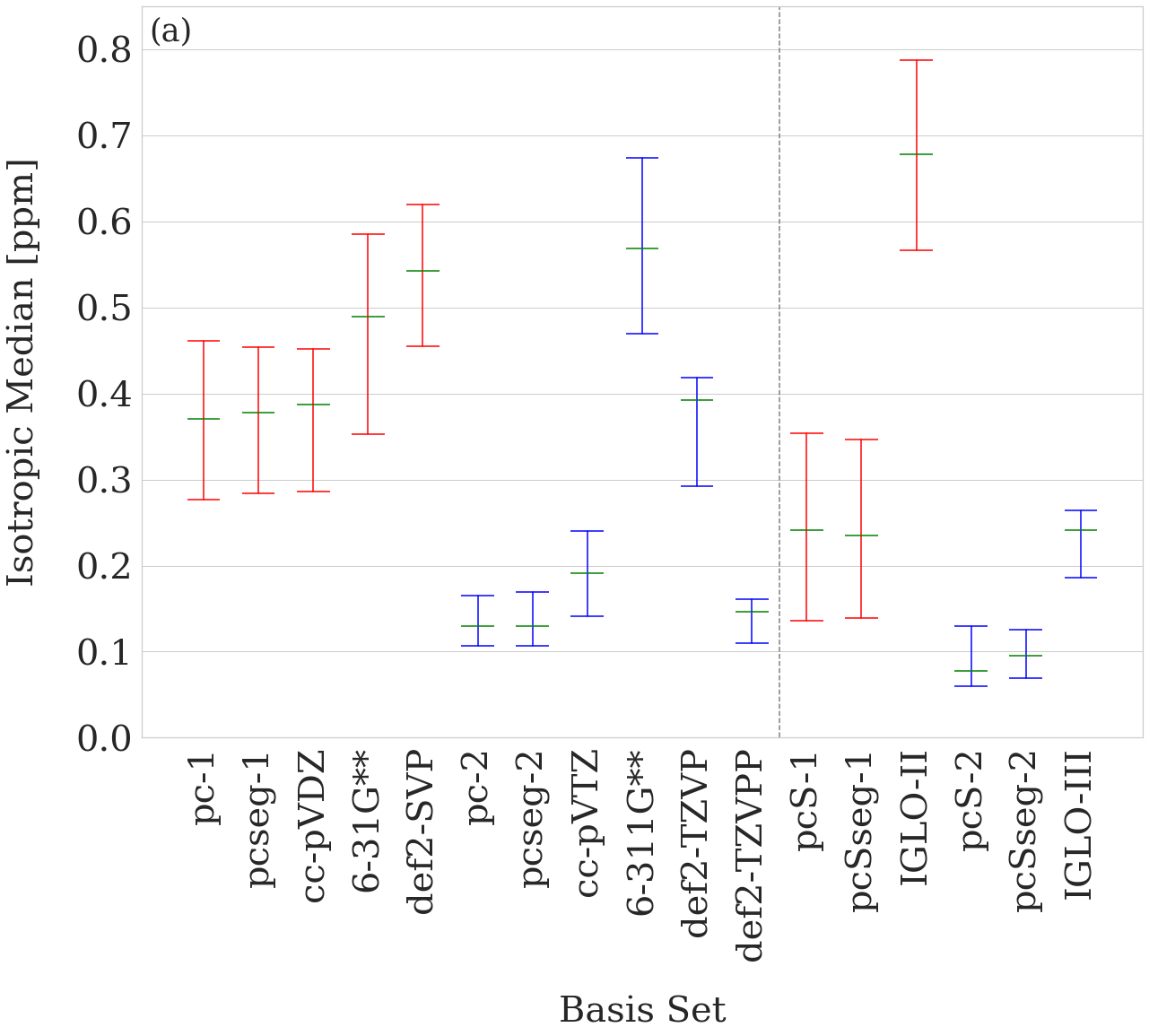}
    \includegraphics[scale=0.2]{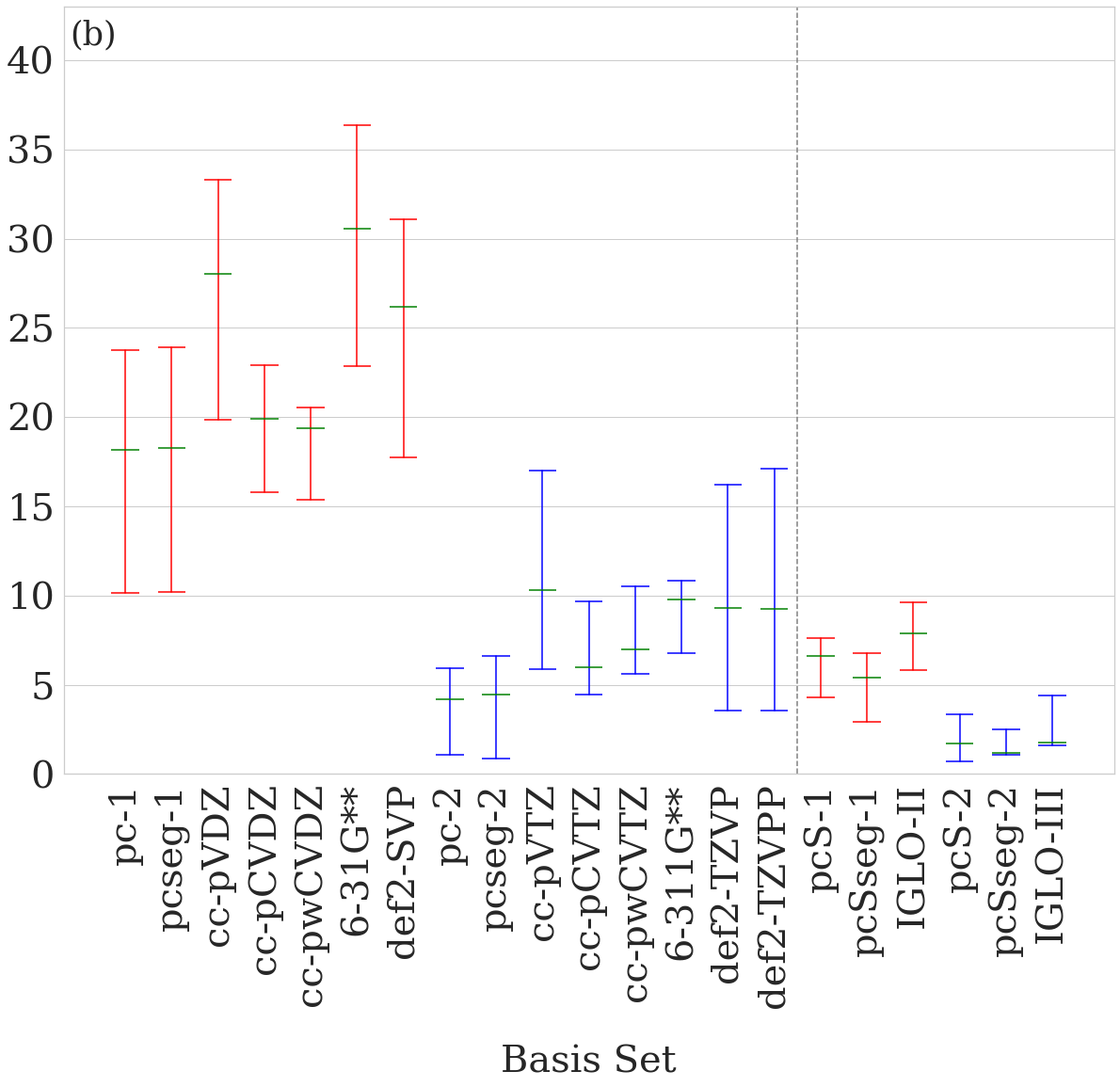}
    \caption{The median error ranges (MERs) for isotropic magnetic shielding constants for (a) hydrogen and (b) non-hydrogen elements. Basis sets to the left of the horizontal grey line are general-purpose basis sets, whilst those to the right are S-specialised basis sets. Ranges in red present double-zeta basis sets, whilst those in blue represent triple-zeta basis sets. For each basis, the median values of the MERs are given as the green horizontal lines.}
    \label{fig:ChemshieldMedianErrors}
\end{figure*}

The absence of the FC term in chemical shielding means that basis sets specialised for these properties do not require tight $s$-primitives or decontraction of $s$-functions in the core region. This is reflected in Figs. \ref{fig:HydrogenFigs} and \ref{fig:CarbonFigs}, where general purpose and S-specialised basis sets have similar $\sD$ and $\smax$. This also explains why the basis set demands for S-specialised basis sets are somewhat different to those of J-specialised and H-specialised basis sets.

Jensen \cite{Jensen2006} found that accurate chemical shielding could be obtained by introducing a tight $p$-function and a slight decontraction, a much more modest change than for the J-specialised and H-specialised basis sets. This is reflected in the $\pmax$ and $\pD$ for the pcS-$n$ and pcSseg-$n$ basis sets in Figs. \ref{fig:HydrogenFigs} and \ref{fig:CarbonFigs}.  The $r^{-3}$ term in the DS operator of the magnetic shielding tensor thus makes the same basis set demands as the $r^{-3}$ term in the PSO operator of the J coupling tensor. Indeed, Jensen found that the PSO term was also improved by the addition of a tight $p$-function. This consistency means that the basis set demands of other properties based on similar operators should be readily predictable from those considered in the three core-properties considered in this paper. 

The IGLO-II and IGLO-III basis sets, both specified for calculation of magnetic shielding constants, are designed differently to the Jensen sets, though still with the variation in the $p$-functions rather than the $s$-functions. Specifically, compared to the Jensen S-specialised basis sets, the IGLO basis sets have  lower $\pmax$ and smaller $\pD$. This reflects a recurring theme that basis sets designed with less contraction have lower exponent Gaussians, implying there is a trade-off between contraction and maximum exponent.

\subsection{Basis Set Performance} \label{Sec:ShieldPerformance}

\begin{table*}[htbp!]
\centering
\caption{Timings for the basis sets used in the calculation of isotropic magnetic shielding constants. Timings are reported relative to those of the 6-31G** basis set (underlined). The S-specialised basis sets are shown in bold. Contraction schemes, as well as the number of basis functions per atom ($N_{\ur{A}}$), for the first row elements are also shown. These timings should be considered as indicative only.}
\begin{tabular}{@{}l c c c c c @{}}
\toprule &
Double-Zeta \hspace{0.2cm} & Triple-Zeta \hspace{0.2cm} &  Rel. Timing & Contraction Scheme & $N_{\ur{A}}$  \\
\midrule
         & def2-SVP  &                   &               0.9 & $3s2p1d$ & 14 \\
         & pcseg-1   &                   &               1.0 & $3s2p1d$ & 14 \\
         & \underline{6-31G**}   &                   &               \underline{1.0} & $\underline{3s2p1d}$ & \underline{14} \\
         & \textbf{pcSseg-1}  &                   &               \textbf{1.1} & $\boldsymbol{3s3p1d}$ & \textbf{17} \\
         & cc-pVDZ   &                   &               1.1 & $3s2p1d$ & 14 \\
         &   \textbf{pcS-1}   &                   &               \textbf{1.1} & $\boldsymbol{3s3p1d}$ & \textbf{17} \\
         &           & 6-311G** &               1.2 & $4s3p1d$ & 18 \\
         & \textbf{IGLO-II}   &                   &               \textbf{1.2} & $\boldsymbol{5s4p1d}$ & \textbf{22}\\
         & cc-pCVDZ  &                   &               1.2 & $4s3p1d$ & 18 \\
         & cc-pwCVDZ &                   &               1.2 & $4s3p1d$ & 18 \\
         &    pc-1   &                   &               1.3 & $ 3s2p1d$ & 14 \\
         &           &         def2-TZVP &               2.0 & $5s3p2d1f$ & 31\\
         &           &          \textbf{IGLO-III} &               \textbf{2.2} & $\boldsymbol{7s6p2d}$ & \textbf{35}\\
         &           &           cc-pVTZ &               2.3 & $4s3p2d1f$ & 30 \\
         &           &        def2-TZVPP &               2.3  & $5s3p2d1f$ & 31 \\
         &           &           pcseg-2 &               2.4 & $4s3p2d1f$ & 30 \\
         &           &              pc-2 &               2.5 & $4s3p2d1f$ & 30 \\
         &           &          \textbf{pcSseg-2} &               \textbf{2.6} & $\boldsymbol{4s5p2d1f}$ & \textbf{36}\\
         &           &             \textbf{pcS-2} &               \textbf{2.7} & $\boldsymbol{4s4p2d1f}$ & \textbf{33} \\
         &           &          cc-pCVTZ &               2.8 & $6s5p3d1f$ & 43 \\
         &           &         cc-pwCVTZ &               2.9 & $6s5p3d1f$ & 43 \\
\bottomrule \label{Table:ChemShieldingsTimings}
\end{tabular}
\end{table*} 

The median error ranges (MERs) and median MER values for isotropic shielding constants are shown for hydrogen and non-hydrogen elements in Figs. \ref{fig:ChemshieldMedianErrors} (a) and (b), respectively.

Immediately we see that, in general, triple-zeta sets perform significantly better than double-zeta sets. There is definitely an improvement in performance for S-specialised vs general-purpose basis sets, particularly for non-hydrogen atoms, but overall there is a smaller improvement when using a specialised basis set than for J coupling and hyperfine property predictions. This is consistent with the fact that the absence of the FC term means that the basis set requirements for shieldings are far more modest than for the other two core-dependent properties.

We recall that the specialised basis sets studied here have two distinct designs; those with larger $\pmax$ and $\pD$ (the Jensen S-specialised basis sets), and those with smaller $\pmax$ and $\pD$ (the IGLO S-specialised basis sets). It is clear from Fig. \ref{fig:ChemshieldMedianErrors} that the former of these two approaches is clearly preferred, especially for hydrogen. The $\pmax$ value of the IGLO-III set is on the same order of magnitude of that of pcS-2, likely leading to their similar errors for the non-hydrogen shielding constants. However, the set-back of the IGLO-II and IGLO-III sets likely comes from an under-represented polarisation space due to the lack of polarisation functions. A well-described polarisation space is likely important due to the pertubative expansion. More details relating performance to the design of basis sets can be found in the Supporting Information.

Table \ref{Table:ChemShieldingsTimings} considers timings relative to 6-31G**. The increase in computational time from specialising basis sets for chemical shielding is more modest than for J coupling constants or hyperfine coupling constants, consistent with the reduced basis set demands for this property. 

The modest increases in computational time in combination with reliable improvements in performance (especially notable for non-hydrogen atoms) means that when predicting magnetic shielding constants, users should use the S-specialised Jensen basis sets, pcSseg-1 (for faster calculations) or pcSseg-2 (for higher accuracy).

\section{Recommendations for Users} \label{sec:Recommendations}

\begin{table}[]
    \centering
    \small
    \caption{Recommended basis sets for J coupling (J), hyperfine (H) and magnetic shielding constants (S)}
    \label{tab:recommendations}
    \begin{tabular}{>{\em}lccccccccccccc}
    \toprule
    & J & H & S \\
    \midrule
\mc{4}{l}{\textbf{Faster calculation}}\\
Basis Set & pcJ-1 & EPR-II & pcSseg-1\\
Rel. time to 6-31G** & 1.8 & 1.6 & 1.1 \\

Median error H & 0.23 Hz & 0.49 MHz & 0.24 pmm \\
Median error X & 0.13 Hz & 1.53 MHz & 5.42 pmm \\
Median error X:H & 0.19 Hz & N/A & N/A \\
    \midrule
\mc{4}{l}{\textbf{Higher accuracy}} \\
Basis Set & pcJ-2 & EPR-III & pcSseg-2 \\
Rel. time to 6-31G** & 6.6 & 4.2 & 2.6 \\
Median error H & 0.05 Hz & 0.70 MHz & 0.10 ppm \\
Median error X & 0.07 Hz & 0.74 MHz & 1.19 ppm \\
Median error X:H & 0.10 Hz & N/A & N/A \\
        \bottomrule
    \end{tabular} 
\end{table}

Our recommended basis sets for the calculation J coupling constants, hyperfine coupling constants and magnetic shielding constants (for chemical shifts) are provided in Table \ref{tab:recommendations}. To summarise Table \ref{tab:recommendations}, the Jensen pcJ-$n$ and pcSseg-$n$ basis sets are recommended for the calculation of J coupling constants and magnetic shielding constants, respectively, whilst the EPR-$n$ bases are recommended for the calculation of hyperfine coupling constants. {The absolute isotropic error distributions for each property are shown in the Supplementary Information for the recommended basis sets.} We note the important caveats that; (a) the benchmark sets are relatively modest in size; (b) timings are indicative only and relative to 6-31G** for the same property.

\section{Final Remarks} \label{Sec:FinalRemarks}
 
In this benchmark study, a number of core-specialised all-Gaussian basis sets are studied, with their performances compared to a number of general-purpose all-Gaussian basis sets with respect to three core-dependent properties. It has been shown that, compared to general-purpose basis sets, core-specialised basis sets are typically characterised by low contraction of the $s$- and $p$-spaces - quantified through the degree of contraction $\iD$ - as well as larger maximum $s$- and $p$-Gaussian exponents ($\imax$ values). For the calculation of three isotropic core-dependent properties, J coupling constants, hyperfine coupling constants and magnetic shielding constants, it was found that basis sets specialised towards each of these properties typically report lower errors than their general-purpose counterparts. 

The study of J coupling constants provides interesting insight into basis set design, given the number of operators with different radial dependencies involved in the calculation of the J coupling tensor. Furthermore, the pertubative expansion with which the majority of the terms are calculated requires excited-state wave functions, giving further relevance to regions further from the nuclear position. The best performing sets were found to have large $s$-exponents, a modest amount of $s$-space contraction and well-described $p$- and polarisation-spaces. These provide the best balance in describing the different terms associated with the final J coupling constant. Of the specialised basis sets investigated, pcJ-1 is the most suitable for fast and accurate calculations, whilst pcJ-2 provides enhanced accuracy at a reasonable increase in computational expense. 

For the calculation of isotropic hyperfine coupling constants, consistently-low errors were found to be achieved by basis sets featuring large $s$-exponents and very little, if any, $s$-space contraction. The similarity of these features to the best performing J-specialised basis sets shows that predictions can be made as to the ideal basis set design by considering the form of the operators involved in calculating the property of interest. For calculating hyperfine coupling constants, the EPR-II basis is recommended for faster calculations whilst the EPR-III basis set provides the best balance between enhanced accuracy and additional computational cost.

For the calculation of isotropic shielding constants, sets containing tight $p$-functions and low $\pD$ values give the lowest errors and most consistent results, especially in the case of hydrogen shielding constants. For faster calculations, the pcSseg-1 basis is recommended, whilst the modest basis set demands of S-specialised basis sets mean that the pcSseg-2 basis can be used for enhance accuracy with a modest increase in computational cost.

Overall, should specialised sets exist for a property, then their use is recommended over general-purpose basis sets. Using different basis sets depending on the property of interest may be seen as somewhat of an inconvenience, even for those well versed in the subtleties of quantum chemistry calculations. This is especially true given the number of basis sets available for use. One would therefore be forgiven for selecting a basis purely on the basis of popularity, especially given that lesser-known specialised basis sets can go unnoticed in such a saturated field. However, just as one should not select a density functional approximation purely on the basis of popularity \cite{Goerigk2019}, the popularity of a particular basis set cannot be the sole justification for its use. We hope to have provided some clarity to this issue, showing that specialised basis sets - should they exist for the property of interest - are likely to be the best option over general-purpose basis sets. 

\section{Acknowledgments}

The authors would like to thank Samuel J. Pitman for insightful comments on the manuscript.

This research was undertaken with the assistance of resources and services from the National Computational Infrastructure (NCI), which is supported by the Australian Government.

The authors declare no conflicts of interest.

\section{Data Availability Statement}

The data underlying this study are available in the online Supporting Information files. 

\section{Supporting Information}
For the three properties and levels of theory investigated, we provide csv files containing: 
\begin{itemize}
    \item Raw results for all constants
    \item Benchmark results for all constants
    \item Absolute isotropic errors for all constants
    \item Median errors for each basis set 
\end{itemize}
We also provide timings for each molecule using the $\omega$B97X-D3 functional. Optimised geometries for all systems at the $\omega$B97M-V/pc-2 level are also provided.
Furthermore, we provide a PDF file describing further analysis on the design of specialised basis sets towards the core-dependent properties discussed.

\bibliographystyle{ieeetr}
\bibliography{BasisSet_LiteratureReview}

\begin{thebibliography}{100}
\bibitem{Notea}
Despite the cc-pV$X$Z, cc-pCV$X$Z and cc-pwCV$X$Z sets being identical for
  hydrogen, differences are observed between these sets for the H:H couplings.
  This suggests an influence from other atoms in the molecule, though this
  analysis is outwith the remit of this paper.
\end{thebibliography}

\end{document}